\RequirePackage{fix-cm}
\documentclass{article}
\usepackage[dvips]{graphicx}
\usepackage{epstopdf}
\usepackage{subfigure}
\usepackage{color}
\usepackage{soul}
\sethlcolor{yellow}
\usepackage{natbib} 
\usepackage{bm}
\usepackage[pagewise]{lineno}
\usepackage{mathptmx}      
\usepackage{amsmath,amssymb,amsfonts}
\usepackage{authblk}
\usepackage[margin=30mm]{geometry}
\newcommand*{\reviewerA}[1]{\textcolor{black}{#1}}
\newcommand*{\reviewerB}[1]{\textcolor{black}{#1}}
\newcommand*{\reviewerC}[1]{\textcolor{black}{#1}}

\newcommand{\argmax}{\mathop{\rm arg~max}\limits}

\begin{document}
\title{Proof-of-concept Study of Sparse Processing Particle Image Velocimetry for Real Time Flow Observation}



\author[1]{Naoki Kanda}
\author[1]{Chihaya Abe}
\author[1]{Shintaro Goto}
\author[1]{Keigo Yamada}
\author[2,1]{Kumi Nakai\footnote{Corresponding to kumi.nakai@aist.go.jp}}
\author[1]{Yuji Saito}
\author[1]{Keisuke Asai}
\author[1]{Taku Nonomura}
\affil[1]{Graduate School of Engineering, Tohoku University}
\affil[2]{National Institute of Advanced Industrial Science and Technology}




\maketitle

\renewcommand{\thefootnote}{}
\footnote[0]{This paper has been accepted for publication in Experiments in Fluids. \copyright 2022. This manuscript version is made available under the CC-BY-NC-ND 4.0 license (https://creativecommons.org/licenses/by-nc-nd/4.0/)}

\begin{abstract}
In this paper, we overview, evaluate, and demonstrate the sparse processing particle image velocimetry (SPPIV) as a real-time flow field estimation method using the particle image velocimetry (PIV), whereas SPPIV was previously proposed with its feasibility study and its real-time demonstration is conducted for the first time in this study.  
In the wind tunnel test, the PIV measurement and real-time measurement using SPPIV were conducted for the flow velocity field around the NACA0015 airfoil model. The off-line analysis results of the test show that the flow velocity field can be estimated from a small number of processing points by applying SPPIV, and also illustrates the following characteristics of SPPIV.  The estimation accuracy improves as the number of processing points increases, whereas the processing time per step increases in proportion to the number of processing points. Therefore, it is necessary to set an optimal number of processing points. Furthermore, the number of modes should be appropriately selected depending on the number of processing points. In addition, the application of the Kalman filter significantly improves the estimation accuracy with a small number of processing points while suppressing the processing time.
When the flow velocity fields with different angles of attack are used as the training data with that of test data, the estimation using SPPIV is found to be reasonable if the difference in angle of attack between the training and test data is equal to or less than 2 deg and the flow phenomena of the training data are similar to that of the test data. For this reason, training data should be prepared at least every 4 deg. 
Finally, the demonstration of SPPIV as a real-time flow observation was conducted for the first time. In this demonstration, the real-time measurement is found to be possible at a sampling rate of 2000 Hz at 20 or less processing points in the top 10 modes estimation as expected by the off-line analyses. 
\end{abstract}

\section{Introduction}
\label{Introduction}
In recent years, active control devices such as plasma actuators have been attracting attention for improving the performance of aircraft \citep{corke2007sdbd, fujii2014high, aono2017plasma, komuro2019influence,sato2020unified}. The performance is expected to be improved by applying feedback control in fluid control using active control devices . \reviewerB{Research efforts to achieve a feedback control have recently increased. For the feedback control of flow separation around an airfoil using the plasma actuator, \citet{benard2011benefits} and \citet{post2006separation} have experimentally investigated a feedback control system using a plasma actuator and pressure sensors for mitigating flow separation around an airfoil at fixed angle of attack and a pitching airfoil, respectively. \citet{segawa2016feedback} have developed a feedback control system using a fiber Bragg grating sensor. While a model-based approach that designs a control strategy in advance is adopted in the previous studies, a model-free approach has been recently proposed. \citet{shimomura2020closed} applied machine learning techniques to determine effective output-input functions utilized in a feedback control system. In addition, feedback flow control methods for various flows such as cylinder wake have been intensively investigated \citep{wu2016closed,varon2019adaptive,brito2021experimental}.} 

It is necessary to sequentially observe the state of the flow field in feedback control, but the target flow velocity is high in the case of aircraft, etc., and therefore, a short time estimation of flow fields is required. One of the methods for measuring the flow velocity field is the particle image velocimetry (PIV). This method is excellent because it is not necessary to install sensors in the model and a two-dimensional flow velocity field can be acquired. Because of these reasons, PIV is considered to be useful when examining feedback control in a wind tunnel test. However, it is difficult to apply it as an observer in feedback control since the computational cost for acquisition of the flow velocity field from the particle image is large. 

\reviewerB{Despite the difficulty above in the computational cost, several approaches have been suggested for the real-time PIV measurement \citep{lelong2003image,yu2006real,kreizer2010real}. Recently, \citet{gautier2015real} achieved the implementation of a dense optical flow algorithm on a graphics processor unit to enable the real-time estimations of flow fields. They demonstrated the feasibility of feedback control using the real-time esimimation of flow fields for the separated flow downstream of a backward-facing step in a hydrodynamic channel \citep{gautier2013control,gautier2015frequency,varon2019adaptive}. Besides, the time-resolved flow fields acquired using the method previously proposed have been also used as inputs for data assimilation and predictions of shear flows \citep{giannopoulos2020data,giannopoulos2020prediction}. However, these studies measured the flow fields at relatively low sampling rates (20 to 224 fps) due to the difficulty in the computational cost. The development of low-cost systems capable of computing flow fields in real time at several kilohertz would allow feedback control for flows exhibiting high-frequency behaviors, such as the separated flow around airfoils.} Therefore, further reduction in the computational cost of the real-time PIV measurement is required. It is effective to apply the sparse sensing that estimates the whole fields from a small number of processing points for the reduction in the computational cost. 

A proper orthogonal decomposition (POD) is a method extracting low-dimensional components from multidimensional data \reviewerB{\citep{berkooz1993proper, holmes1996turbulence,lumley1967structure,bonnet2001review,boree2003extended}}. Here, POD is a method of extracting as much energy as possible with the smallest number of bases. When this is applied to fluid analysis, it is possible to extract dominant flow field structures \citep{taira2017modal}. When the flow velocity field can be reduced in dimension by POD etc., the entire flow velocity field can be estimated from the limited processing points using the reconstruction using optimized sparse sensor \citep{manohar2018data,clark2018greedy,saito2021determinant,yamada2021fast,nakai2021effect,carter2021data,li2021efficient}, which is also applied to other flow problems \citep{kaneko2021data,inoue2021data,inoba2022optimization,li2021data,fukami2021global,callaham2019robust}. In this method, it is important to determine the appropriate observation position. A method for locating processing points in two-dimensional data such as a flow velocity field (a vector sensor problem) has been proposed by \citet{joshi2009sensor,saito2020data,saito2021data}. \reviewerB{In addition, the system identification methods of flow fields using POD have been intensively investigated.} 
\citet{suzuki2014pod,nankai2019linear,suzuki2020few,nonomura2021quantitative} proposed a discrete linear model that estimates the time evolution of the flow velocity field constructed from their low-dimensional POD modes. \reviewerB{\citet{inigo2014dynamic,inigo2016recovery} proposed a technique to build a reduced state-space model to predict the dynamics of a flow from local wall measurements using POD or dynamic mode decomposition (DMD) in noise-amplifier flow. Furthermore, system identification techniques based on DMD \citep{nonomura2018dynamic,nonomura2019extended} and those combining POD and neural networks \citep{giannopoulos2020data,giannopoulos2020prediction,deng2019super,hasegawa2020machine,hasegawa2020cnn} have been investigated. }

The present authors applied these techniques to the Kalman filter and proposed sparse processing PIV (SPPIV) as a real-time flow velocity field measurement method of low computational cost \citep{kanda2021feasibility}. \reviewerA{This SPPIV technique significantly reduces the computational costs, and therefore, can be used for the real-time observation. The real-time observation of velocity fields leads to flow control and real-time diagnostics of anomaly in fluid machineries that would be used for from academic study to industrial application. 
Furthermore, the extension of SPPIV for laser imaging, detection, and ranging (LiDER) sensor measurement might be one of practical implementation for future real-time flow-control and diagnostics of flow around aircraft and automobiles, though LiDER measurement has more restrictions than SPPIV has and the further issues should be resolved before its application. Because the SPPIV technique can be used for fundamental academic study and also can be the foundation of the further development of real-time observations, the demonstration of the real-time observation and further investigation of its performance should be clarified.}

\reviewerB{The previous study \citep{kanda2021feasibility} proposed the framework of SPPIV and just discussed about the feasibility of SPPIV in the simplified condition and did not demonstrate the real-time observation using SPPIV. In the present study, further detailed offline analyses are conducted for the evaluation of SPPIV, and then, the real-time observation using SPPIV is demonstrated for the first time.}

\section{Overview of Sparse Processing PIV}
\label{Sparse Processing PIV}
\begin{figure}
    \centering
    \includegraphics[width=11cm]{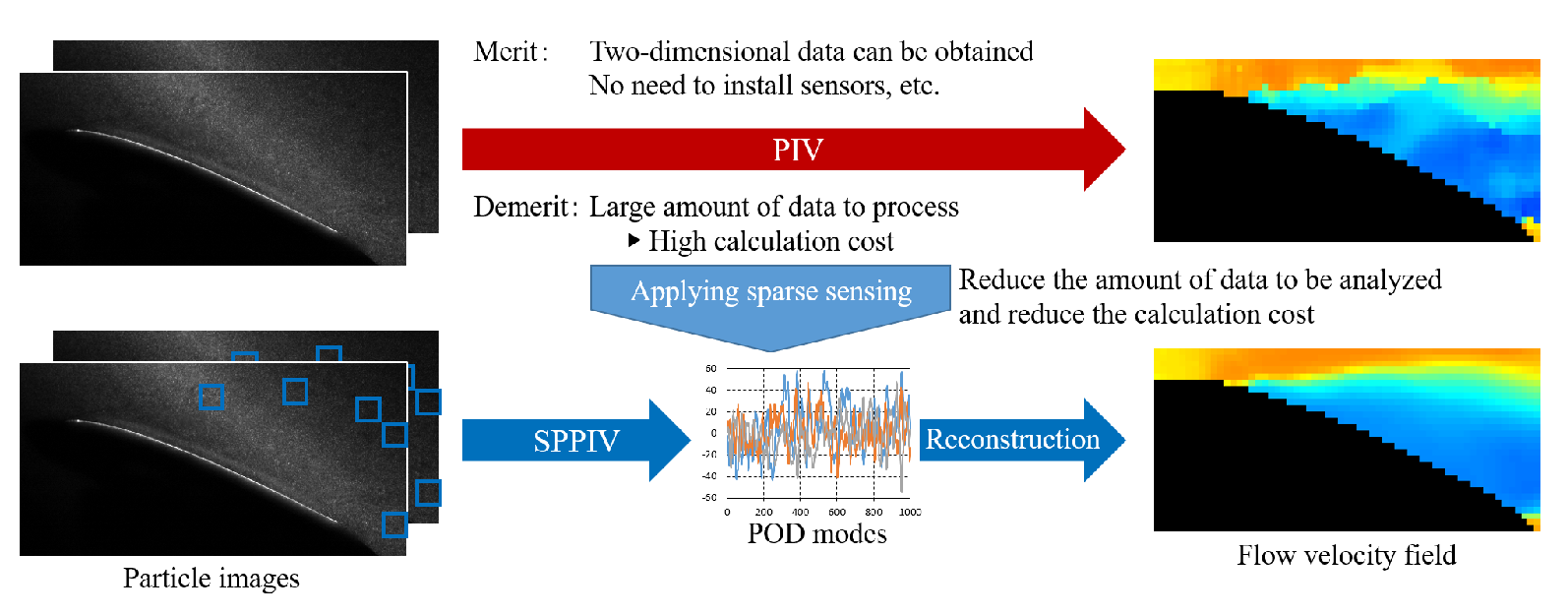}
    \caption{The schematic of SPPIV}
    \label{fig:schem_SPPIV}
\end{figure}

\begin{figure}
    \centering
    \includegraphics[width=10cm]{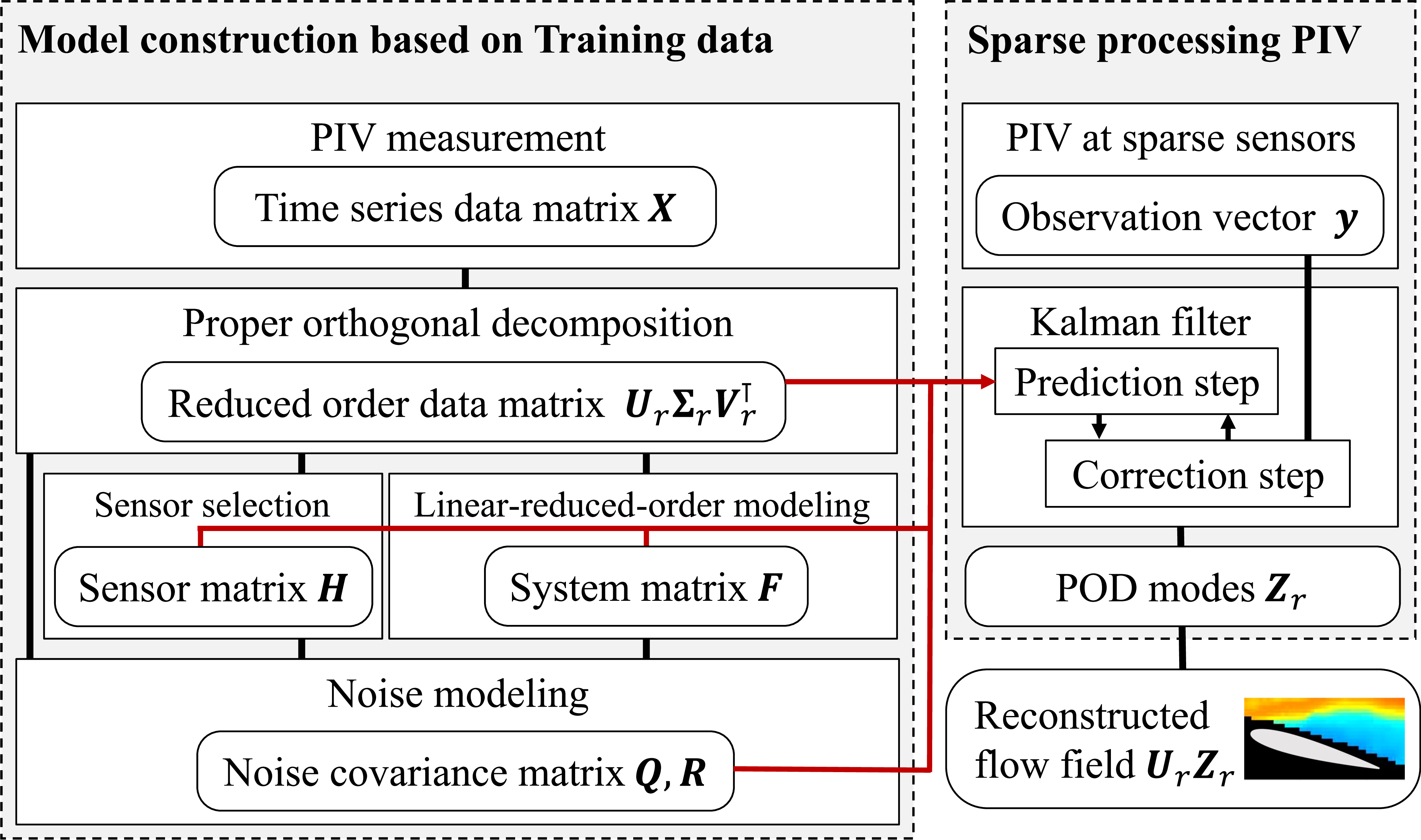}
    \caption{\reviewerB{The flowchart of SPPIV}}
    \label{fig:FC_SPPIV}
\end{figure}

Again, SPPIV is a method of estimating the flow field from the sparse velocity vectors obtained by conducting PIV using a limited number of correlation windows in the particle image as processing points \citep{kanda2021feasibility}. The schematic of SPPIV and the flowchart of this method are shown in Figs.~\ref{fig:schem_SPPIV} and \ref{fig:FC_SPPIV}, respectively. A linear Kalman filter is usually used for SPPIV together with the POD mode representations. Therefore, the POD mode presentation which is a basis of SPPIV is described in Section 2.1. The Kalman filter with the state-space approximation, which is employed in a standard SPPIV implementation as described by \citet{kanda2021feasibility}, is described in Section 2.2. In addition, SPPIV can be realized without using the Kalman filter which is not recommended in the present study, and this implementation is explained in Section 2.3 for the discussion of the performance improvement of use of the Kalman filter.  

\subsection{Proper orthogonal decomposition}
A proper orthogonal decomposition (POD) is one of the low-dimensional methods extracting as much energy as possible from the minimum number of bases \reviewerB{\citep{berkooz1993proper,lumley1967structure,bonnet2001review,boree2003extended, taira2017modal}}.
The basis obtained by POD is called a POD mode. When applying POD to fluid analysis, it is possible to reduce the dimension of flow field data obtained by PIV measurements and extract important flow field phenomena \citep{taira2017modal}.
The formula when POD is applied to the data matrix $\bm{X}$ is shown in Eq.~\ref{eq:POD}. In this research, we constructed a data matrix in which snapshots of time-series data are reshaped into one-dimensional column vectors and are arranged in the row direction. Therefore, $\bm{U}_r$, $\bm{\Sigma}_r$, and $\bm{V}_r$ represent the reduced spatial mode, amplitude, and temporal mode, respectively, and $r$ is the number of modes. The flow velocity field reduced by POD is represented by the superposition of spatial modes weighted by $\bm{\Sigma}_r\bm{V}_r^{\mathsf{T}}$. Therefore, the flow velocity field can be estimated using $\bm{\Sigma}_r\bm{V}_r^{\mathsf{T}}$, if the spatial mode is obtained from training data.
In this study, $\bm{\Sigma}_r\bm{V}_r^{\mathsf{T}}$ will be referred to as $\bm{Z}$ hereafter.
\begin{linenomath}
\begin{align}
\label{eq:POD}
    \bm{X}=\bm{U}\Sigma\bm{V}^{\mathsf{T}}\approx\bm{U}_r\Sigma_r\bm{V}_r^{\mathsf{T}}=\bm{U}_r\bm{Z}.
\end{align}
\end{linenomath}
\reviewerB{
Assume that $\bm{U}_r$ is the constant in time, the dominant structure of the instantaneous entire flow fields vector $\bm{x}_n$ can be reconstructed by the corresponding instantaneous truncated POD mode coefficient vector $\bm{z}_n$. 
\begin{linenomath}
\begin{align}
\label{eq:POD}
    \bm{x}_n \sim \bm{U}_r\bm{z}_n.
\end{align}
\end{linenomath}
Therefore, the following SPPIV framework introduces the methods estimating $\bm{z}_n$.}

\subsection{SPPIV: State estimation using Kalman filter}
\reviewerB{
Here, the standard implementation of SPPIV using the Kalman filter is described. For the Kalman filter implementation, the state-space model for the POD mode coefficient is required. In this section, the state-space model is introduced first, and then, the Kalman filter is explained. After that the data-driven estimation of the state and observation equations and the noise models, which are used for the state-space models, are described. 
}

\subsubsection{Linear state-space model}
The linear state-space model can be described as follows:
\begin{linenomath}
\begin{align}
\label{eq:state}
    \bm{z}_{k+1 \mid k}&=\bm{F} \bm{z}_{k \mid k} + \bm{v}_k,\\
\label{eq:obs}
    \bm{y}_{k}&=\bm{C} \bm{z}_k + \bm{w}_k,
\end{align}
\end{linenomath}
The estimated state vector is $\bm{z}$, the observed value is $\bm{y}$, and the system noise and the observation noise are $\bm{v}_n$ and $\bm{w}_n$, respectively. Then, the estimated value and the observed value are expressed by Eqs.~\ref{eq:state} and \ref{eq:obs}, respectively. \reviewerB{The sparse observation $\bm{y}$ of flow velocity data is obtained by performing PIV at the position determined by the sensor location optimization method. In this study, the flow velocity is calculated by the fast-Fourier-transformed cross-correlation method as the PIV at the processing point position with 32 $\times$ 32 pixel$^2$ interrogation windows. Here, a `processing point' in the framework of SPPIV corresponds to a `sensor' in the framework of the sparse sensor location optimization since the selected sparse calculation is conducted in SPPIV while the measuring device has already installed in the framework of the sparse sensor location optimization without any selection. } 
Here, $\bm{F}$ in Eq.~\ref{eq:state} and $\bm{C}$ in Eq.~\ref{eq:obs} are the system and  observation matrices, respectively. The deviations of those matrices are in Section. 2.2.3. 

\subsubsection{Kalman filter: a main module of SPPIV}
The Kalman filter is an algorithm that sequentially calculates the least squares estimate of the state vector based on the noise-disturbed observations in a linear system in which the observation and system noise follow a Gaussian distribution \citep {kalman1960new}. This method consists of a prediction step and a filter step. In the prediction step, the next state is predicted from the data up to the previous step using a linear model. In the filter step, the current state is estimated from the current observation value disturbed by noise and the predicted value obtained in the prediction step.

Here, the prediction step in the Kalman filter is expressed by \reviewerB{Eqs.~\ref{eq:upstate} and \ref{eq:upcov}}. In \reviewerB{Eqs.~\ref{eq:upstate} and \ref{eq:upcov}}, $\hat{\bm{z}}$ represents the state vector estimated by the Kalman filter, and the system matrix $\bm{F}$ is obtained from the training data with assuming the time-invariant system, as later discussed in Section \ref{sec:lrom}.
\begin{linenomath}
\begin{align}
\label{eq:upstate}
    \hat{\bm{z}}_{k+1 \mid k}&=\bm{F} \hat{\bm{z}}_{k \mid k},\\
\label{eq:upcov}
    \bm{P}_{k+1 \mid k}&=\bm{F} \bm{P}_{k \mid k}\bm{F}^{\mathsf{T}} + \bm{Q},
\end{align}
\end{linenomath}
The filter step is expressed by \reviewerB{Eqs.~\ref{eq:kalmangain}, \ref{eq:obsstate} and \ref{eq:obscov}}. 
\begin{linenomath}
\begin{align}
\label{eq:kalmangain}
    \bm{K}_k&=\bm{P}_{k \mid k-1}\bm{C}^{\mathsf{T}}(\bm{C}_k\bm{P}_{k \mid k-1}\bm{C}^{\mathsf{T}}+\bm{R})^{-1},\\
\label{eq:obsstate}
    \hat{\bm{z}}_{k \mid k}&=\hat{\bm{z}}_{k \mid k-1}+\bm{K}_k(\bm{y}_k-\bm{C}\hat{\bm{z}}_{k \mid k-1}),\\
\label{eq:obscov}
    \bm{P}_{k \mid k}&=\bm{P}_{k \mid k-1}-\bm{K}_k\bm{C}\bm{P}_{k \mid k-1},
\end{align}
\end{linenomath}
where $\bm{P}$ and $\bm{K}$ represent the error covariance matrix and Kalman gain, respectively, and $\bm{Q}$ and $\bm{R}$ represents a noise covariance matrix for the system and observation, respectively.  For any matrix $\bm{A}$, $\bm{A}_{n|n-1}$ is the $n$th estimated data from the up to the ($n-1$)th data, and $\bm{A}_ {n|n}$ is estimated $\bm{A}$ using up to $n$th data.
In this study, the initial values of the error covariance matrix $\bm{P}$ and the estimated state vector $z$ are the identity matrix and the zero vector, respectively. The matrices $\bm{F}$, $\bm{C}$, $\bm{Q}$, and $\bm{R}$ are constructed in the data-driven procedure as described in the following subsections.

\subsubsection{Data-driven construction of model}
The framework of SPPIV stands on the data-driven estimation of flow fields, and therefore, the system identification and the model construction based on the training data is required. 
The time-series training data matrix of the state values $\bm{z}$ and the observed values $\bm{y}$ are defined as shown in Eqs.~\ref{eq:Z_matrix} and \ref{eq:Y_matrix}, respectively. Also, when using \reviewerB{the first to $r$th modes} for estimation, $\bm{\Sigma}_r$ and $\bm{V}_r$ are lower-dimensional amplitude and temporal POD modes, respectively. $\bm{H}$ in Eq.~\ref{eq:Y_matrix} represents the sensor location matrix determined by the sensor location optimization method described later in this section. Here, $N$ is the number of time-series data in Eqs.~\ref{eq:Z_matrix} and \ref{eq:Y_matrix}.

\begin{linenomath}
\begin{align}
\label{eq:Z_matrix}
    \bm{Z}&=[\bm{z}_1 \ \bm{z}_2 \cdots \bm{z}_N]=\bm{\Sigma}_r\bm{V}_r^\mathsf{T},\\
\label{eq:Y_matrix}
    \bm{Y}&=[\bm{y}_1 \ \bm{y}_2 \cdots \bm{y}_N]=\bm{H}\bm{X},
\end{align}
\end{linenomath}

\paragraph{Linear-reduced-order-model-based system equation}
\label{sec:lrom}
In this study, the low-dimensional discrete linear model proposed by \citet{nankai2019linear} is applied, and the system matrix $\bm{F}$ in Eq.~\ref{eq:upstate} is estimated with assuming $\bm{Z}$ as the state variables. This method identifies the coefficient matrix $\bm{F}$ by linearly approximating the time variation of POD modes $\bm{Z}$. In this method, two matrices are obtained from POD mode $\bm{Z}$ using Eqs.~\ref{eq:Zn-1} and \ref{eq:Zn}. The system matrix $\bm{F}$ can be obtained by applying Eq.~\ref{eq:sysmatrix} to these equations. Here, $(\bm{Z}_{m-1})^{+}$ represents the Moore-Penrose pseudoinverse of $\bm{Z}_{m-1}$.
\begin{eqnarray}
\label{eq:Zn-1}
    \bm{Z}_{m-1}&=&[\bm{z}_1 \ \bm{z}_2 \cdots \bm{z}_{m-2} \ \bm{z}_{m-1}],\\
\label{eq:Zn}    
    \bm{Z}_{m}&=&[\bm{z}_2 \  \bm{z}_3 \cdots \bm{z}_{m-1} \ \bm{z}_m],\\
\label{eq:sysmatrix}
    \bm{F}&=&\bm{Z}_m(\bm{Z}_{m-1})^{+}.
\end{eqnarray}
\reviewerA{
This estimation approach is actually identical to DMD-model-based estimations. One of them was conducted by \citet{gomez2019data} in which a DMD model is used with a Kalman filter for flow estimation with the pressure sensors as observation.   Equation~\ref{eq:sysmatrix} is actually identical to the equation of the exact DMD matrix in the projected POD subspace, which is the standard practice for deriving the $\bm{F}$ matrix in exact DMD. It should be noted that the difference between exact DMD and standard DMD in subspace is just in an order of processes of \reviewerC{SVD \citep{brunton2019data}.}
Additionally, \cite{berry2017application,berry2017dmd} conducted series of studies on off-line estimation of flow fields using POD models as the present author did in the present study.  
}

\paragraph{Processing point optimization and observation equation}
\label{processing point position optimization}
Thus far, several sensor optimization methods have been proposed by \cite{joshi2009sensor,manohar2018data,manohar2018optimal,manohar2019optimized,clark2018greedy,clark2020multi,clark2020sensor,yamada2021fast,nonomura2021randomized,nakai2021effect,nakai2022nondominated,nagata2022randomized,nagata2022data,li2021data,li2021efficient,inoue2022data}. 
In this study, the greedy optimization method for the sensor location in the vector field proposed by \citet{saito2021data} is used as the standard optimization method of processing points. In this method,  $\bm{X}=[\bm{X}_u^{\mathsf{T}}~\bm{X}_v^{\mathsf{T}}]^{\mathsf{T}}$ is assumed where $\bm{X}_u^{\mathsf{T}}$ and $\bm{X}_v^{\mathsf{T}}$ represent the data matrices of the freestream- and vertical-direction components of the velocity field. The spatial mode $\bm{U} = [\bm{U}_u^{\mathsf{T}}~\bm{U}_v^{\mathsf{T}}]^{\mathsf{T}}$ is obtained by POD applied to data matrix $\bm{X}$ whereas $\bm{U}_u^{\mathsf{T}}$ and $\bm{U}_v^{\mathsf{T}}$ represent the spatial modes of freestream and vertical direction components of the velocity field.
\reviewerA{In this case, the observation equation with $k$ sparse sensors can be written as follows:
\begin{eqnarray}
\bm{y}&=&\left[\begin{array}{c}
\bm{I}_2 \otimes \bm{h}_{i_{1}} \\
\bm{I}_2 \otimes \bm{h}_{i_{2}} \\
\vdots \\
\bm{I}_2 \otimes \bm{h}_{i_{k}} \\
\end{array}\right]
\left[\begin{array}{c}\bm{U}_u \\ \bm{U}_v \end{array}\right]\bm{z}
\nonumber
     = \left[\begin{array}{c}\bm{H}_{s_{1}}\nonumber\\
                               \bm{H}_{s_{2}}\nonumber\\
                               \vdots\\
                               \bm{H}_{s_{k}}\\\end{array}\right]
                               \left[\begin{array}{c}\bm{U}_u \\ \bm{U}_v   \end{array}\right] \bm{z}
     = \left[\begin{array}{c}\bm{W}_{1}\\
                               \bm{W}_{2}\\
                               \vdots\\
                               \bm{W}_{k}\\\end{array}\right]\bm{z}\nonumber\\
     &=& \bm{C}_k \bm{z}, 
\end{eqnarray}
where $\bm{I}_2$ represents the $2\times2$ identical matrix, $\bm{h}_{i_k}$ the $k$th sensor location row vector which has unity on the sensor location index $i_k$ and zeros on the others, $\bm{H}_{s_{k}}=\bm{I}_2 \otimes \bm{h}_{i_k}$, $\bm{W}_k = \bm{H}_{s_{k}} \left[\begin{array}{cc}\bm{U}_u^{\mathsf{T}} & \bm{U}_v^{\mathsf{T}} \end{array}\right]^\mathsf{T}$ and 
$\bm{C}_k=\left[\begin{array}{cccc}\bm{W}_1^{\mathsf{T}} & \bm{W}_2^{\mathsf{T}} & \dots & \bm{W}_k^{\mathsf{T}} \end{array}\right]^\mathsf{T} $. 
The D-optimal design of experiments selects the position with which the determinant of the Fisher information matrix represented by Eq.~\ref{eq:Fisher} becomes larger as the sensor location, and the corresponding $\bm{C}_k$ matrix is constructed:
\begin{eqnarray}
\label{eq:Fisher}
    \textrm{maximize} \left\{
    \begin{array}{l}
    \left|\bm{C}_k\bm{C}_k^{\mathsf{T}} \right| ~(sp \leq r) \\
    \left|\bm{C}_k^{\mathsf{T}}\bm{C}_k \right| ~(sp > r)
    \end{array}
    \right.,
\end{eqnarray}
where $r$, $s$ and $p$, represents the number of modes considered for the system, the number of components of vector sensor which is 2 in this study, and the number of vector sensors, respectively. However, optimization of this index is formulated to be a \reviewerC{combinatorial problem} and is nonpolynomial-time hard. Therefore, a greedy heuristic method is employed and the objective function is suboptimized. In the greedy method, $i_k$ is selected and $\bm{C}_k$ is maximized for given $\bm{C}_{k-1}$ in each incremental step.
The greedy method extended to the vector sensors of multiple components at one point was employed, since the observed value used in this study is the flow velocity vector obtained by PIV which has two components in the freestream and vertical directions at one processing point. It is possible to select positions suitable for reconstructing the entire flow field with the minimum estimation error by applying this method.
Here, the formulation of the greedy method for vector sensors is only shown without derivations.
\begin{eqnarray}
\label{eq:Fisher}
i_{k}=\argmax_{i\, \in\, \mathcal{S}\, \backslash\, \mathcal{S}_{k-1}} 
\left\{ 
\begin{array}{ll} \det \left( \bm{W}_i \left(  \bm{I}-\bm{C}_{k-1}^{\mathrm{T}} 
\left(\bm{C}_{k-1}\bm{C}_{k-1}^{\mathrm{T}}\right)^{-1}\bm{C}_{k-1}
\right) \bm{W}_i^{\mathrm{T}} 
\right) 
& \quad sk \le r \\
\det\left(\bm{I} + \bm{W}_i\left(\bm{C}_{k-1}^{\mathrm{T}}\bm{C}_{k-1}\right)^{-1} \bm{W}_i^{\mathrm{T}}\right). & \quad sk > r 
\end{array} 
\right.,
\end{eqnarray}
where $\mathcal{S}$ and $\mathcal{S}_k$ represents the sets of all the possible sensors and of selected sensors up to the $k$th incremental step. See reference \citep{saito2021data} for the details of the derivation.  The objective function above for the sensor location of each incremental step can be derived after the cumbersome calculation described in reference \citep{saito2021data}.
}

Assuming that the sensor location matrix obtained by this method is $\bm{H}$ of which a component of a selected sensor location is unity and others are zeros, the observation equation expressing the relationship between the observation value $\bm{y}$ and the state vector $\bm{z}$ is represented by Eq.~\ref{eq:y_cz}. 
\begin{eqnarray}
\label{eq:y_cz}
    \bm{y}=\bm{H}_s\bm{U}_{r}\bm{z}=\bm{C}\bm{z},
\end{eqnarray}
where $\bm{C}_k $ is described to be $\bm{C}$ for brevity. 
See reference \citep{saito2020data,saito2021data}, for more details of the greedy method and the structure of the $\bm{H}_s$ matrix for vector sensors. 

\paragraph{Noise Model}
In this study, the noise covariance matrices were also calculated from the training data using Eqs.~\ref{eq:state}, \ref{eq:obs}, \ref{eq:errorcov_v} and \ref{eq:errorcov_w}. The off-diagonal components of $\bm{Q}$ and $\bm{R}$ are assumed to be 0. Also, $\bm{Q}$ and $\bm{R}$ are obtained from the training data and are assumed to be time-invariant in the present study.
\begin{eqnarray}
\label{eq:errorcov_v}
    {Q_{ij}}&=&\frac{1}{m-1}\sum_{k=1}^{m-1}{v}_{k_{i}}{v}_{k_{j}} \delta_{ij},\\
\label{eq:errorcov_w}
    {R_{ij}}&=&\frac{1}{m}\sum_{k=1}^{m}{w}_{k_{i}}{w}_{k_{j}} \delta_{ij},
\end{eqnarray}
where, $Q_{ij}$ and $R_{ij}$ denote the $(i,j)$-component of $\bm{Q}$ and $\bm{R}$, respectively, $v_{k_i}$ and $w_{k_i}$ denote the $i$th-component of $\bm{v}_k$ and $\bm{w}_k$, respectively. Here, $\delta$ denotes the Kronecker delta.
$\bm{v}_k$ and $\bm{w}_k$ of training data are calculated from substituting Eq.~\ref{eq:sysmatrix} into Eqs.~\ref{eq:state} and \ref{eq:obs}, respectively.

\subsection{Another Implementation of SPPIV: Pseudo Inverse Implementation}
The SPPIV can be conducted using snapshot-to-snapshot sparse sensor reconstruction which has been introduced by previous studies \citep{manohar2018data,saito2021determinant,saito2021data}. 
In this implementation, the mode amplitude can be reconstructed by instantaneous observation, and the time advancement is no longer considered. The observation equation Eq.~\ref{eq:obs} is only employed, and the estimation is given by using the pseudoinverse matrix of $\bm{C}$. 
\begin{eqnarray}
\label{eq:est_woKF}
    \bm{z}_k&=\bm{C}^{+} \bm{y}_{k}.
\end{eqnarray}
The readers should refer the reference \citep{saito2021data} for the details of this specific formulation because the vector sensors are adopted unlike the original works by \cite{manohar2018data,saito2021determinant}. We do not recommend this implementation for the SPPIV with fewer processing points in terms of accuracy, as discussed in Section \ref{sec:WOKF}.

\section{Wind Tunnel test}
\label{Wind Tunnel test}
\subsection{Experimental equipment}
\label{Experimental equipment}
\begin{figure}
    \centering
    \includegraphics[width=8cm]{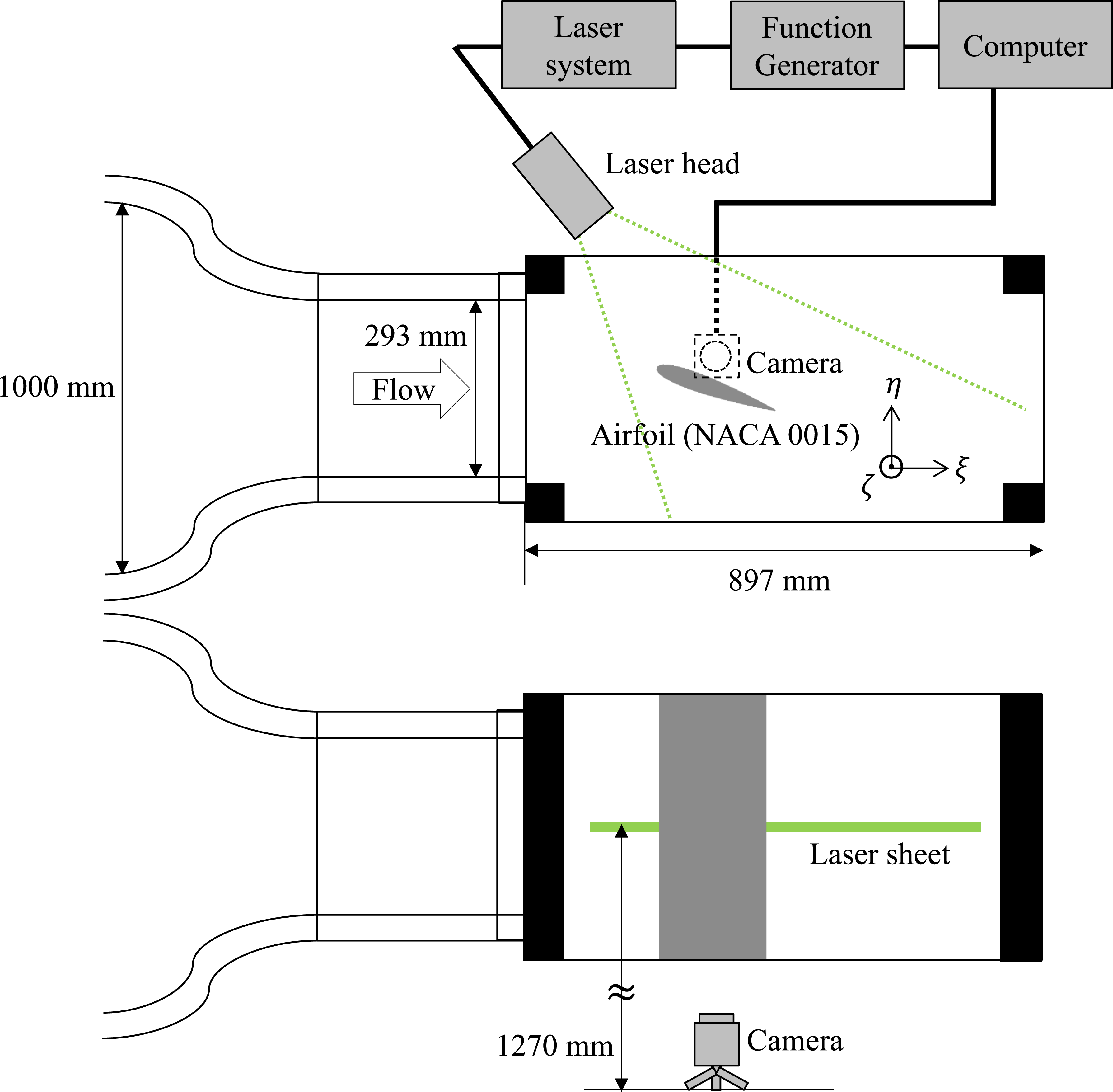}
    \caption{\reviewerB{The setup of wind-tunnel testing in top and side views. Here, the wind tunnel contour is not scaled.}}
    \label{fig:ExpSetup}
\end{figure}

The experiment was conducted in the Small Low-Turbulence Wind Tunnel at the Institute of Fluid Science, Tohoku University. The wind tunnel is a single-way circulation type wind tunnel, and an open type test section was used in this study. The outlet of the contraction nozzle of the wind tunnel has a regular octagonal cross section with the distance ($D$) of 0.293~m from one side to the other. Figure \ref{fig:ExpSetup} shows the setup of the experiment. \reviewerB{The turbulence intensity is less than 0.03~\%, and the profile of the turbulence intensity is uniform in the range of -0.25 $< \eta/D <$ 0.25 and -0.25 $< \zeta/D <$ 0.25, which corresponds to the region excluding the boundary layer near the wall.} The airfoil model of NACA0015 was employed in the experiment. The chord length and span width of this model are 100~mm and 300~mm, respectively. This model is made by stereolithography and has 29 static pressure holes on the upper and lower surfaces of the airfoil. The angle of attack is set to be 0 deg so that the static pressure distributions on the upper and lower surfaces of the symmetric airfoil match each other when the flow is given. Although, in this study, an error of approximately 0.2 deg to 0.5 deg in the angle of attack is included due to the influence of the experimental setup when adjusting 0 deg, this error does not alter the conclusion of the paper which discusses the evaluation and demonstration of SPPIV.

The PIV measurement system used in this study consists of a high-speed camera (IDP-Express R2000, Photron), a double-pulse Nd:YLF laser (DM30-527-DH, Photonics Industries), and a function generator (WF1948, NF). The lens used was a Nikkor 35mm f/1.4 manufactured by Nikon. Here, the high-speed camera used in the experiment can directly \reviewerB{transfer} the image data onto the computer memory, and the image data are subsequently processed by computer. This camera is one of the key component of the real-time observation. This camera has been used for dynamic projection mapping \citep{narita2016dynamic} or other applications. \reviewerB{The size of the sensor of this camera is 5.12 $\times$ 5.12~mm$^2$. The images of resolution of 512 $\times$ 512~pixel$^2$ (corresponding to the full frame) can be acquired at a sampling rate of 2000~fps, and the images of resolution of 512 $\times$ 96~pixel$^2$ can be acquired at a maximum sampling rate of 10,000~fps. In this study, the sampling rate of 4000~fps of images (corresponding to the sampling rate of 2000~Hz of the velocity field) and the image resolution of 512 $\times$ 256~pixel$^2$ were employed. Note that the sampling rate of 4000~fps is the maximum for the camera with the image resolution of 512 $\times$ 256~pixel$^2$. With regard to the laser, the power is 45~W per each individual head, and the maximum pulse frequency is 10~kHz. The laser interval was $t = 80 \mathrm{\mu}$s} \reviewerA{with the frame straddling technique.}
Figure~\ref{fig:ExpSetup} shows that the laser light source was installed on the starboard upstream side of the wind tunnel test section, and the laser sheet was irradiated in the direction parallel to the floor surface. Dioctyl sebacate droplets generated by a Ruskin nozzle was used as the tracer particles.

\subsection{Test condition}
\label{Test Condition}
In the experiment, the freestream velocity $U_\infty$ was set to 10~m/s. The corresponding \reviewerB{chord} Reynolds number was $Re \approx 6.7 \times 10^4$. \reviewerB{In PIV measurement, the interrogation window of 32 $\times$ 32~pixel$^2$ and overlap of 75~\% were employed, leading to the spatial resolution of 0.2517~mm the number of vectors/field of 29 $\times$ 61. Note that the number of vectors/field excluding the mask covering the airfoil is 819. The size of the interrogation window is an important parameter to affect the spatial resolution of the flow velocity field. The influence of the size of the interrogation window will be investigated in our future works.} 
\reviewerB{The angle of attack was set to be $\alpha = 14, 16, 18, 20, 22$~deg for the offline analyses and only 18~deg for the real-time demonstration.} When SPPIV is incorporated into the control system as an observer in the future study, the control will be conducted for the separated flow to be attached. The transition from the attached flow to the separated flow, which is of interest in the present and future studies, occurs near the condition of the minimum angle of attack of 14~deg in these measurement conditions. When the real-time measurement was conducted for an angle of attack $\alpha = 18$~deg, the measurement parameters were set to be the exactly same as those of training data. 

\subsection{Computational condition}
Table~\ref{table:Computational enviroment} shows the specifications of the computer used in this study. In the experiment, time-series data of the flow velocity field is acquired  by PIV measurement. The test conditions are the same as the test data, as described in Section \ref{Test Condition}. The sensor location matrix $\bm{H}$ and noise covariance matrices $\bm{Q}$ and $\bm{R}$ used for the training data were calculated by the in-house code written in MATLAB.

\begin{table}[htbp]
    \centering
        \caption{The specification of the computer}
        \begin{tabular}[t]{ll}
        \hline\hline
        Processor information&Intel(R) Xeon(R) W2265 CPU\\
        \reviewerB{Processor base frequency} & \reviewerB{3.5 GHz}\\
        \reviewerB{The number of cores} & \reviewerB{12}\\
        Random access memory&64.0 GB\\
        System type&64-bit operating system\\
        &x64 base processor\\
        \hline\hline
        \label{table:Computational enviroment}
    \end{tabular}
\end{table}

In this study, an offline analysis was conducted using the particle images obtained in the wind tunnel test and the performance of SPPIV was investigated. In the offline analysis, the effects of three different parameters on the change in the performance of SPPIV were investigated. 
Firstly, the effects of the number of processing points $p$ and the number of modes $r$ on the estimation accuracy and the processing time per step are investigated. Secondly, 
the estimation accuracy is investigated when only using the observation equation with that of SPPIV without the dynamical system. Finally, the estimation accuracy is investigated for the case in which the flow velocity fields with different angles of attack from test data are used as training data. 
The estimation accuracy was evaluated by 5-fold cross-validation, except for the investigation of the estimation accuracy when the flow velocity fields with different angles of attack were used as the training data. Cross-validation was conducted using five sets of 1000 pairs of particle images obtained in this test.
When the training data with a different angle of attack from the test data was used, the training data of 3000 pairs of particle images were obtained in this test.
Table~\ref{table:SPPIV_offline_param} shows the settings of various parameters in these analysis cases.
Regarding the processing time per step of SPPIV in the offline analysis, the time required to process one pair of particle images was obtained by averaging 1000 steps in one test, and the average processing time and the standard deviation of the results of the 100 tests are calculated.

\begin{table}[htbp]
    \centering
    \caption{Experimental parameters in offline analysis}
    \begin{tabular}[t]{cccccc}
    \hline\hline
    Test case& Number of modes & Number of processing points & angle of attack\\
    & $r$ & $p$ & $\alpha$ (deg) \\
    \hline
    Effect of $r$ and $p$ & 10, 20, 30 & 5 - 100 & 18  \\
    & 40, 50 & &  \\
    \hline
    Effect of estimation & 10, 20, 30 & 5 - 819 & 18 \\
    w/ and w/o Kalman filter & 40, 50 & (measuring time: 5 - 100)&\\
    \hline
    Effect of using training data & 10 & 5, 10, 15, & 14, 16, 18, \\
    with different angle of attack &  & 20, 25, 30 & 20, 22 \\
    \hline\hline
    \label{table:SPPIV_offline_param}
    \end{tabular}
\end{table}

\begin{figure}
    \centering
    \includegraphics[width=11cm]{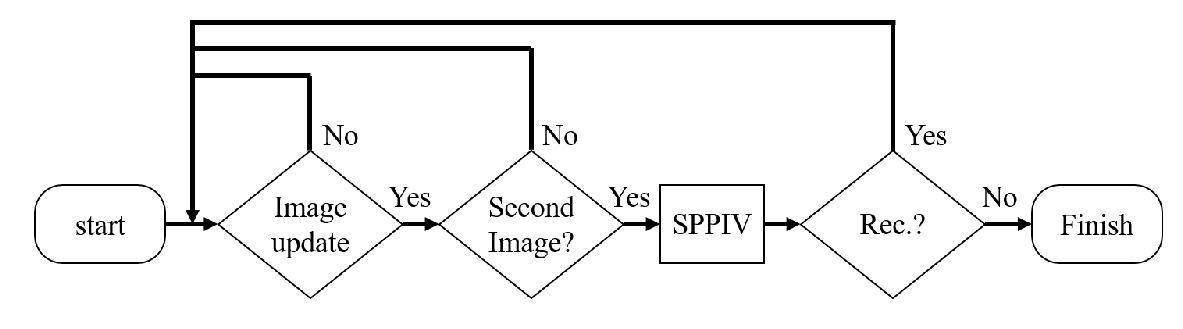}
    \caption{The flowchart of real-time measurement}
    \label{fig:Flowchart_realtime}
\end{figure}

In the real-time measurement, the number of modes estimated was set to be 10, and the number of processing points was set to be $p =$ 5, 10, 15, 20, 25, and 30. The development environment and languages used for real-time measurement programs are Microsoft Visual Studio 2019 and C++, respectively. \reviewerA{The software development kit for the real-time camera control provided by Photron was employed.} In the real-time measurement, 3830 particle images could be acquired, but the first image cannot be used due to the synchronization between the laser and the start of recording with the camera and computer, and thus, the second to 3829th particle images (1914 pairs of particle images) are used. Here, the flowchart of the real-time measurement is shown in Fig.~\ref{fig:Flowchart_realtime}. 

\reviewerA{In the real-time measurement, images are taken and saved by the camera regardless of the execution of SPPIV. Image data are streaming on the memory of computer, and the selected point of interrogation windows are only processed by program written by C++. Therefore, in determining the success of the real-time measurement, the counted number of images taken and the number of images for which SPPIV was conducted were compared.} 
The condition in which the numbers are corresponding to each other was considered to be the necessary condition for the success of the real-time measurement. In addition, since the judgment based only on the counted numbers is not sufficient, the success or failure of the real-time measurement should be judged in the present state by comprehensively considering the estimation accuracy, the result of the offline analysis using the acquired image, and the processing time. 
\reviewerA{For example, we use a frame straddling technique with naive implementation of the program code, and the first and second images in pairs are fixed. Once the first and second images are exchanged by delay of processing, the error becomes significantly large in the present naive implementation of the program.} This point should be further discussed in the future study.

\section{Offline Parametric Study}
This section describes the results of an offline parametric study for the purpose of evaluating the characteristics of SPPIV for real-time measurements.
\reviewerA{It should be stressed that the cross-validation analysis is conducted for all the analyses as described in the previous section.}
\subsection{Effect of number of processing points and modes on error and processing time}
\begin{figure}
    \centering
    \includegraphics[width=11cm]{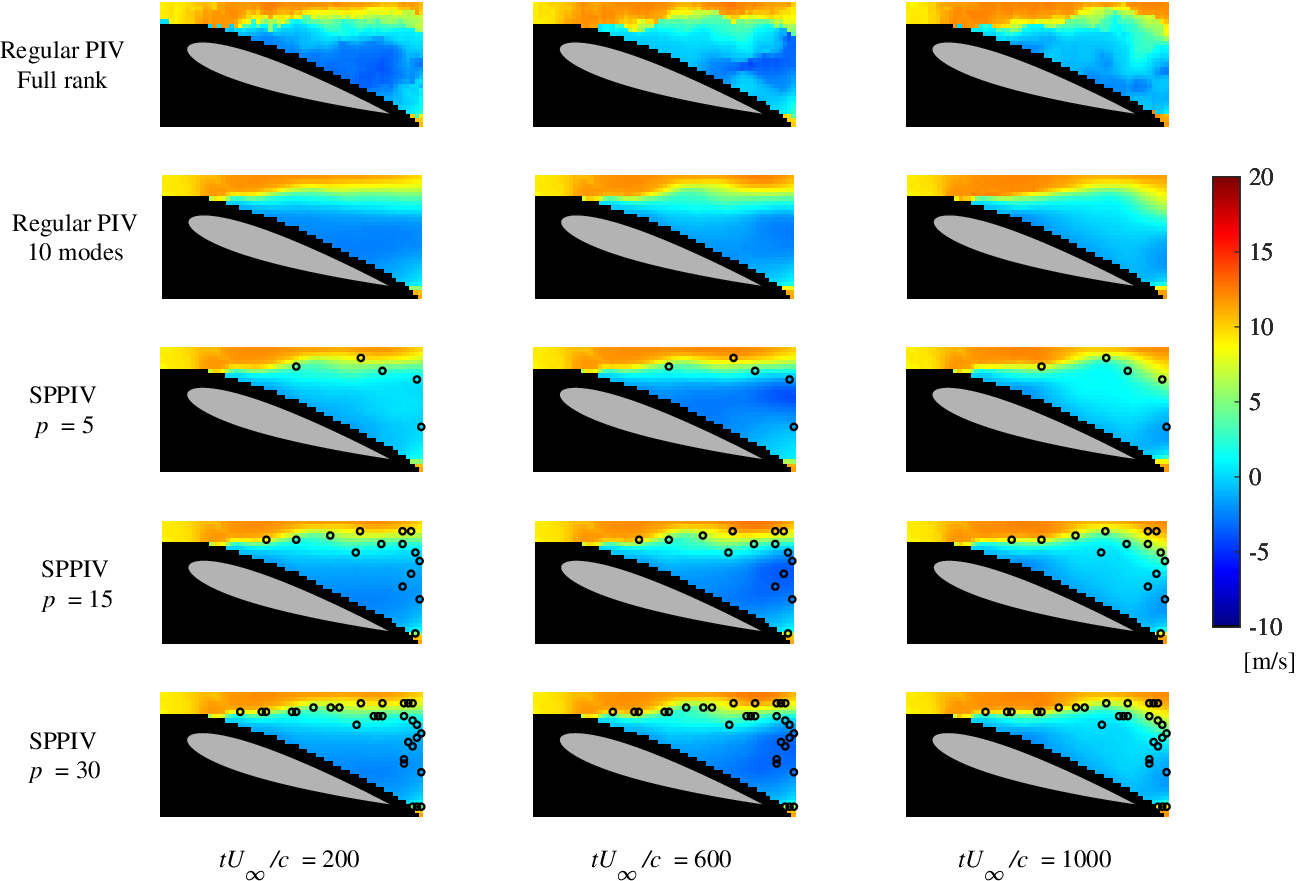}
    \caption{The velocity fields of the freestream direction ($r$ = 10)}
    \label{fig:verocityfieldsu_direction}
\end{figure}

\begin{figure}
    \centering
    \includegraphics[width=12cm]{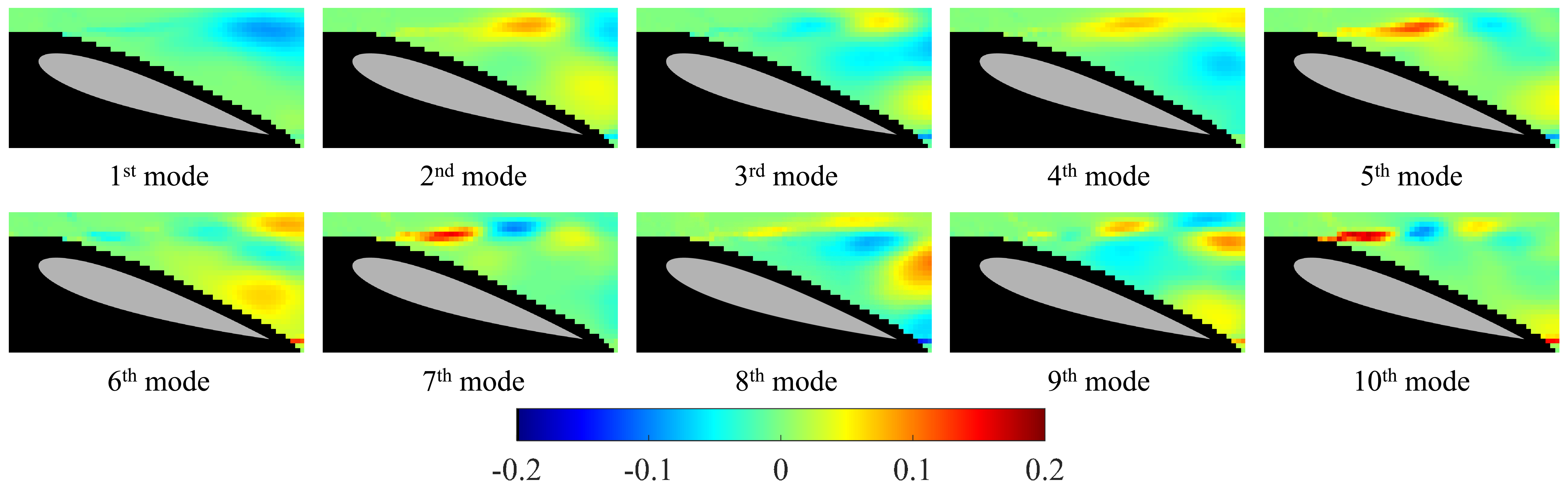}
    \caption{\reviewerB{The distributions of the spatial modes of the first to tenth POD modes ($\alpha$ = 18)}}
    \label{fig:spatial_mode}
\end{figure}

\begin{figure}
    \centering
    \includegraphics[width=6cm]{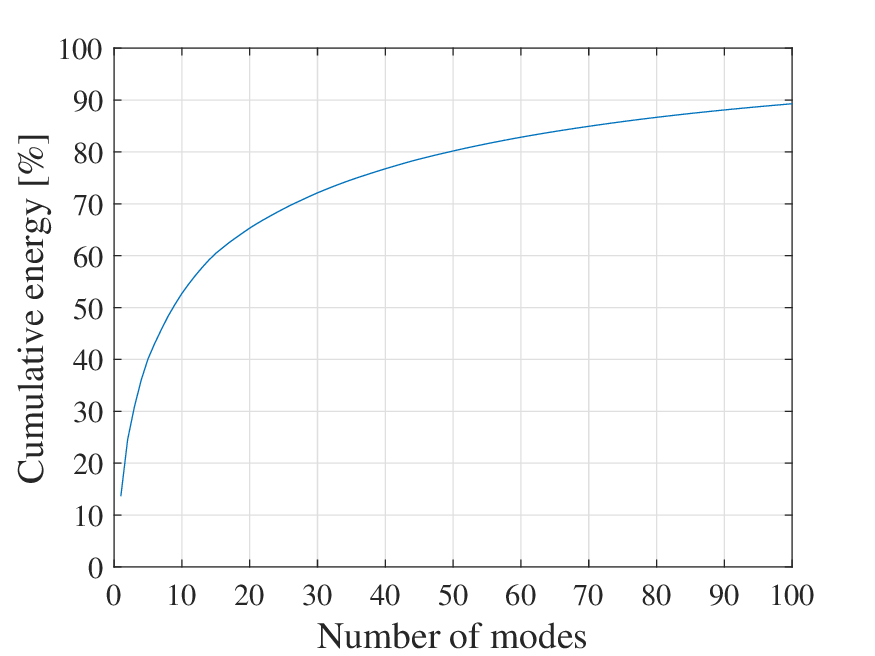}
    \caption{\reviewerB{The relationship between the energy reconstruction ratio and the number of modes used for the dimensional reduction ($\alpha$ = 18)}}
    \label{fig:reconst_ratio}
\end{figure}

First, the cases with a different number of processing points and modes used for estimation (the number of elements of the estimated state vector) were investigated. As an example of the measured flow velocity field, Fig.~\ref{fig:verocityfieldsu_direction} shows the freestream direction component of the velocity field obtained by PIV, that reconstructed using only the top 10 modes, and that estimated by SPPIV. Figure~\ref{fig:verocityfieldsu_direction} illustrates that the low-dimensional data was reconstructed using the spatial mode of the training data and the POD mode obtained by projecting the test data into the spatial mode of the training data using Eq.~\ref{eq:ref_PODmode}. 
\reviewerB{Here, Fig.~\ref{fig:spatial_mode} shows the distributions of the spatial modes of the representative POD modes. These correspond to the spatial modes of training data for one of the five sets of 1000 pairs used for 5-fold cross-validation. Figure~\ref{fig:spatial_mode} illustrates that the representative modes express large fluctuations near the boundary layer between the acceleration and recirculation regions. On the other hand, higher modes was confirmed to express smaller flow structures near the boundary layer and the leading edge and the variation in the recirculation region. 
In addition, Fig.~\ref{fig:reconst_ratio} shows the relationship between the energy reconstruction ratio of dimensional reduction and the number of modes used for the dimensional reduction. The energy reconstruction ratio is calculated by dividing the sum of squares of each component of the amplitude of the low-dimensional data up to $r$ mode ($\bm{\Sigma}_{r}$) by that of the full-rank data ($\bm{\Sigma}$). The reconstruction ratio is 52.7~\% in the case of using the top 10 modes for the dimensional reduction. It indicates that the significant structures of the separated flow field can be expressed in the top 10 modes. Figure~\ref{fig:verocityfieldsu_direction} illustrates the low-dimensional data can reproduce relatively large fluctuations near the boundary layer between the acceleration and recirculation regions.
In this case, qualitatively almost the same results were obtained by applying an independent POD to the test data instead of projecting it to the training data. 
On the other hand, the modes up to tenth are not considered to include the variation of the recirculation region.}

\reviewerB{Figure~\ref{fig:verocityfieldsu_direction} also illustrates that the processing points tend to be located around the shear layer. The spatial distribution of the processing points are determined by the optimization algorithm explained in Section~\ref{processing point position optimization}. Basically, the area with the large fluctuation of the flow are preferentially selected as the processing points using the algorithm. The algorithm selects the points located on the antinodes (in a broad sense) of the spatial modes of each POD mode, while the superposition of multiple modes are evaluated in the selection. Figure~\ref{fig:verocityfieldsu_direction} shows that the antinodes of the spatial modes of the representative POD modes are located around the shear layer. The spatial distribution of the processing points has affect on the estimation accuracy of the flow field. \citet{saito2021data} reported that the estimation accuracy using randomly selected points are lower than that using the points selected by the optimization algorithm. Note it could be difficult to install processing points or sensors at the antinodes of the spatial modes in practical. In such cases, the positions are selected using a constrained sensor optimization algorithm which takes into account the cost of installing processing points or sensors \citep{clark2018greedy}.}

In addition, comparison of the low-dimensional data using the top 10 modes and estimated data using SPPIV shown in  Fig.~\ref{fig:verocityfieldsu_direction} illustrates that it is possible to estimate relatively large fluctuations that can be expressed in the top 10 modes from the information of a small number of processing points. 


\begin{figure}
    \centering
    \includegraphics[width=12cm]{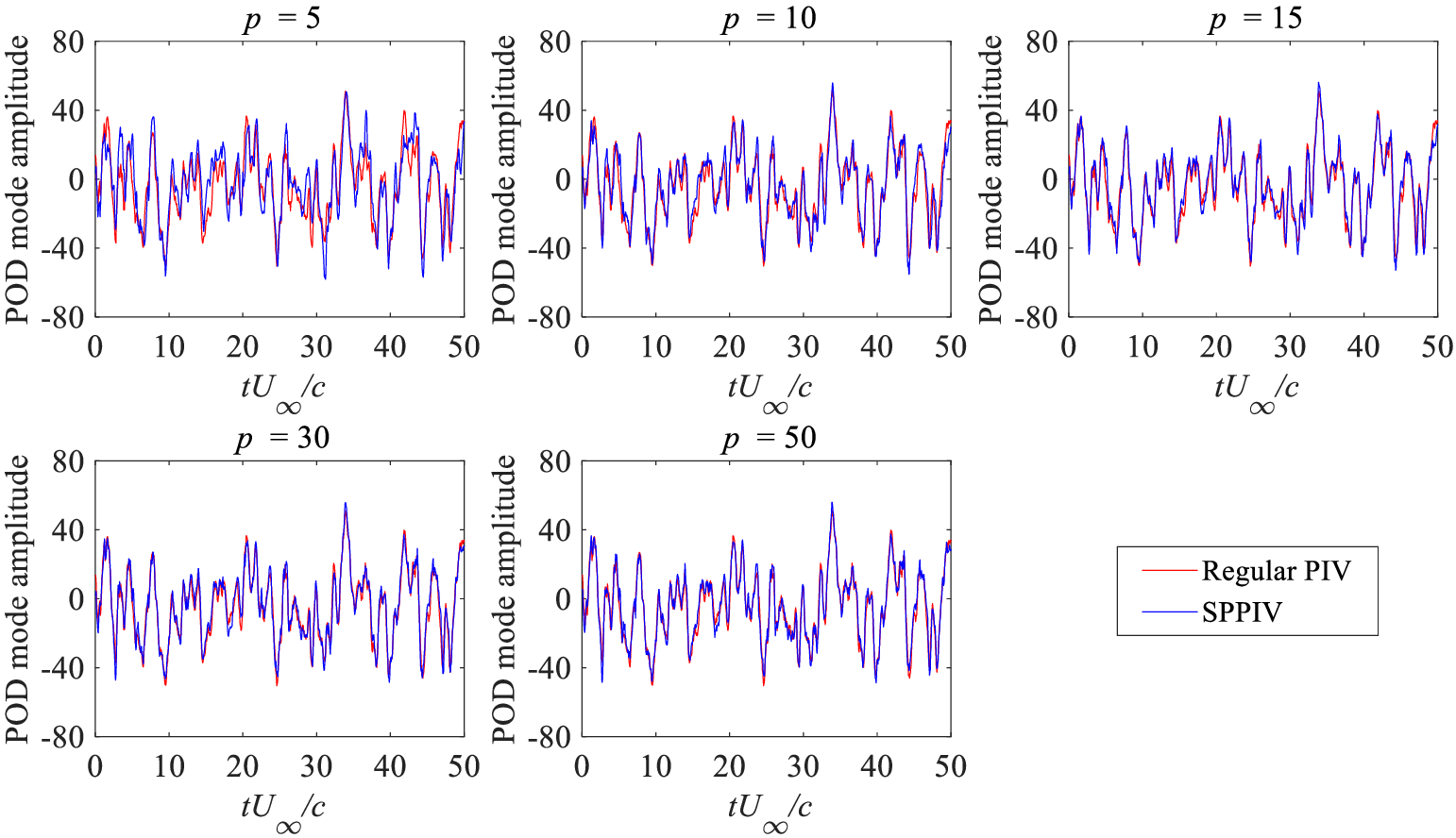}
    \caption{\reviewerB{The time histories of the first mode amplitude obtained using PIV and estimated using SPPIV in the case of $p =$ 5, 10, 15, 30, and 50, and $r =$ 10}}
    \label{fig:timehistoriesofPODmodes_r10}
\end{figure}

\begin{figure}
    \centering
    \includegraphics[width=12cm]{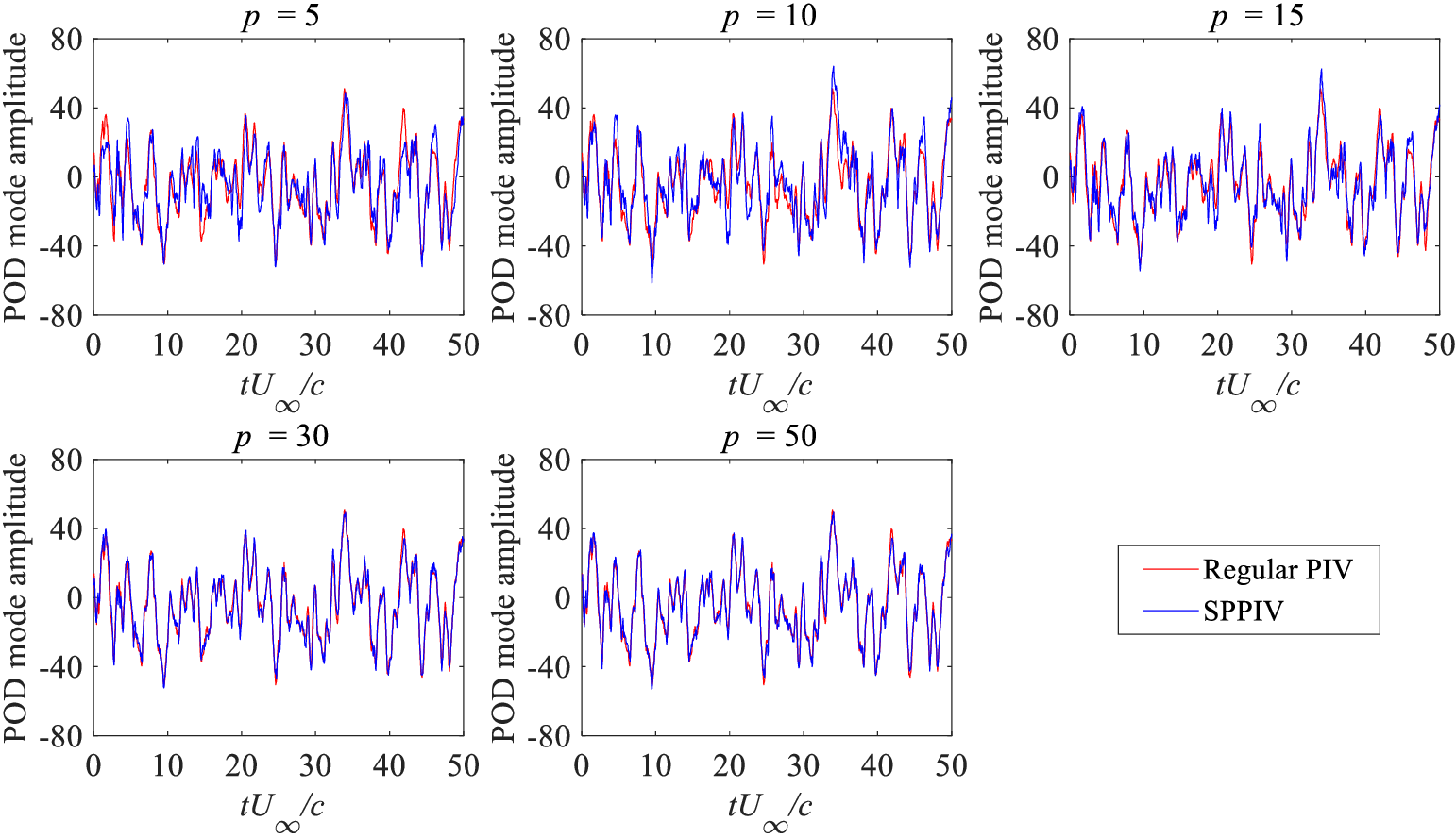}
    \caption{\reviewerB{The time histories of the first mode amplitude obtained using PIV and estimated using SPPIV in the case of $p =$ 5, 10, 15, 30, and 50, and $r =$ 30}}
    \label{fig:timehistoriesofPODmodes_r30}
\end{figure}

\begin{figure}
    \centering
    \includegraphics[width=11.5cm]{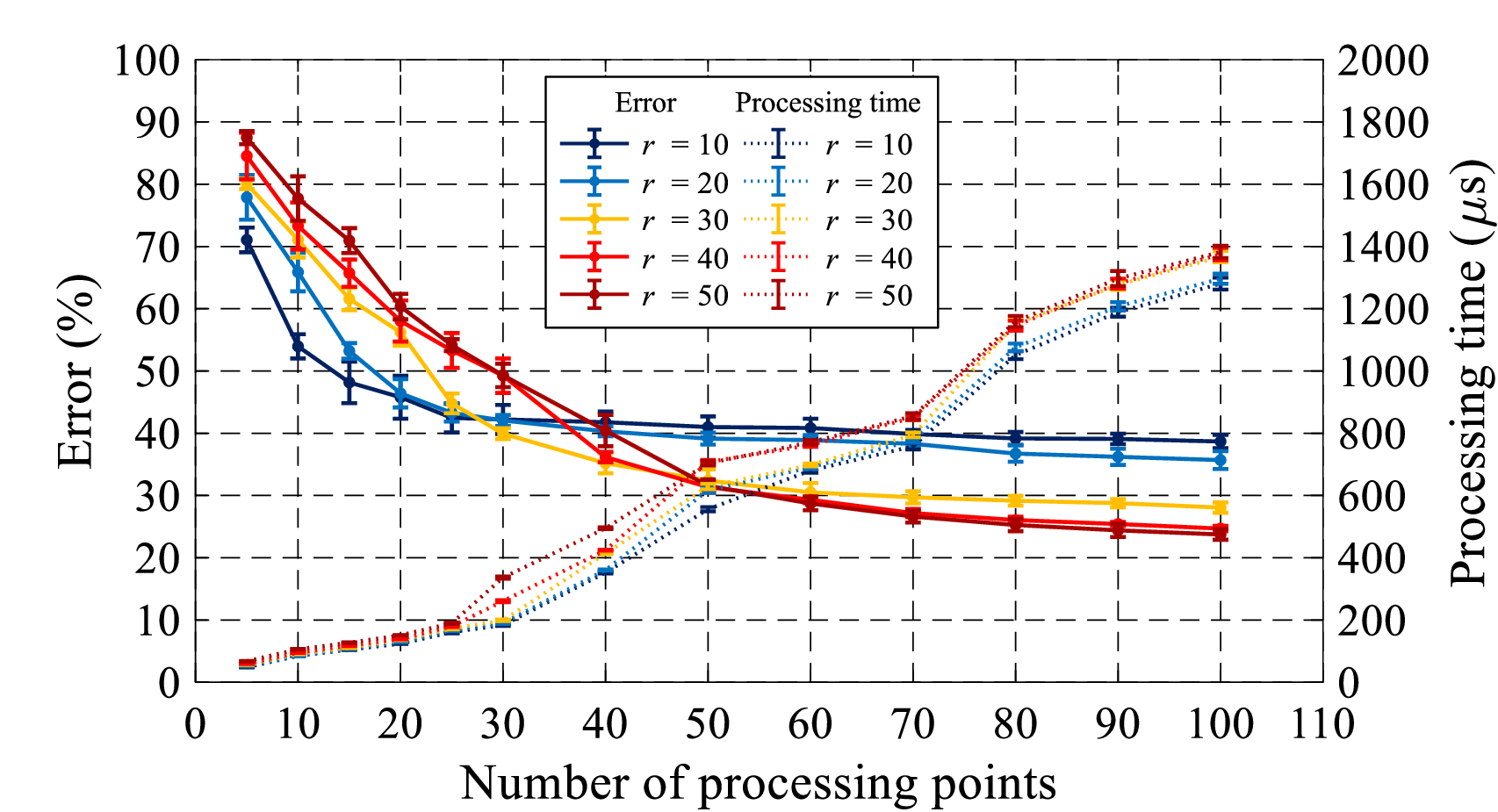}
    \caption{\reviewerB{The error and the processing time of SPPIV}}
    \label{fig:errorofSPPIV}
\end{figure}

Figures~\ref{fig:timehistoriesofPODmodes_r10} and \ref{fig:timehistoriesofPODmodes_r30} show a part of the time histories of the representative POD mode amplitude obtained using PIV and SPPIV in the case of $r=10$ and $r=30$, respectively. The error is calculated using Eq.~\ref{eq:mode_error_all}. The errors when changing the number of processing points and the number of modes are quantitatively evaluated and shown in Fig.~\ref{fig:errorofSPPIV}. Here, $Z^{\textrm{ref}}$ in Eq.~\ref{eq:mode_error_all} is obtained by projecting the data matrix of the time-series flow velocity field obtained by PIV into the spatial mode of the training data used in SPPIV by using Eq.~\ref{eq:ref_PODmode}. 


\begin{linenomath}
\begin{align}
\label{eq:ref_PODmode}
    \bm{Z}^{\rm{ref}} &= \bm{U}_r{^\mathsf{T}}\bm{X}
\end{align}
\end{linenomath}

\begin{eqnarray}
\label{eq:mode_error_all}
    \epsilon &= \frac{\sqrt{\sum_{i=1}^{r}\sum_{j=1}^{m}(Z_{i,j}^{\rm{ref}}-\hat{Z}_{i,j})^2}}{\sqrt{\sum_{i=1}^{r}\sum_{j=1}^{m}{(Z_{i,j}^{\rm{ref}})^2}}}
\end{eqnarray}

Figure~\ref{fig:errorofSPPIV} shows that the estimation accuracy is improved by increasing the number of processing points. \reviewerB{The estimation error is lower in the case of the smaller number of modes when the number of processing points is 30 or less. The comparison of Fig.~\ref{fig:timehistoriesofPODmodes_r10} with Fig.~\ref{fig:timehistoriesofPODmodes_r30} illustrates that when the processing point is 5, 10, and 15, the variation of the first mode amplitude is estimated more accurately in the case of the number of modes of 10 compared to that of 30. This is because the snapshot observation system using the number of modes of 30 is underdetermined when the number of processing points is less than 15, while that of the number of modes of 10 becomes overdetermined when the number of processing points exceeds 5. Thus, the estimation error of the representative POD mode becomes large in the underdetermined condition for the snapshot observation system. Although the estimation errors are lower in the case of the number of modes of 10 to 30 when the number of processing points is less than 30, the estimation error becomes lower in the case of a large number of modes such as 50 when the number of processing points is greater than 40.} The estimation accuracy in higher-order modes is expected to be improved by increasing the number of processing points more than the number of modes. 
%
%
Figure~\ref{fig:errorofSPPIV} also shows that the processing time per step increases in proportion to the increase in the number of processing points. Since the estimation accuracy improves as the number of processing points increases as shown in Fig.~\ref{fig:errorofSPPIV}, there seems to be a trade-off between the estimation accuracy and the processing time. Therefore, it is necessary to set the number of processing points so that the estimation accuracy and the calculation time could be balanced for the real-time measurement with high accuracy.

In addition, the effects of the number of modes are discussed with the same plots in Fig.~\ref{fig:errorofSPPIV} but from the different point of view. 
The errors of the case with the smaller number of modes are smaller when the number of processing points are smaller than 30, while those of the case with larger number of modes are smaller when the number of processing points are larger than 40. These results show that the number of modes should be increased with the number of the processing points. The number of the modes
is implicated to be appropriately selected depending on the number of processing points as shown in Fig.~\ref{fig:errorofSPPIV}.

\subsection{Comparison of estimation with and without Kalman filter}
\label{sec:WOKF}
\begin{figure}
    \centering
    \includegraphics[width=11.5cm]{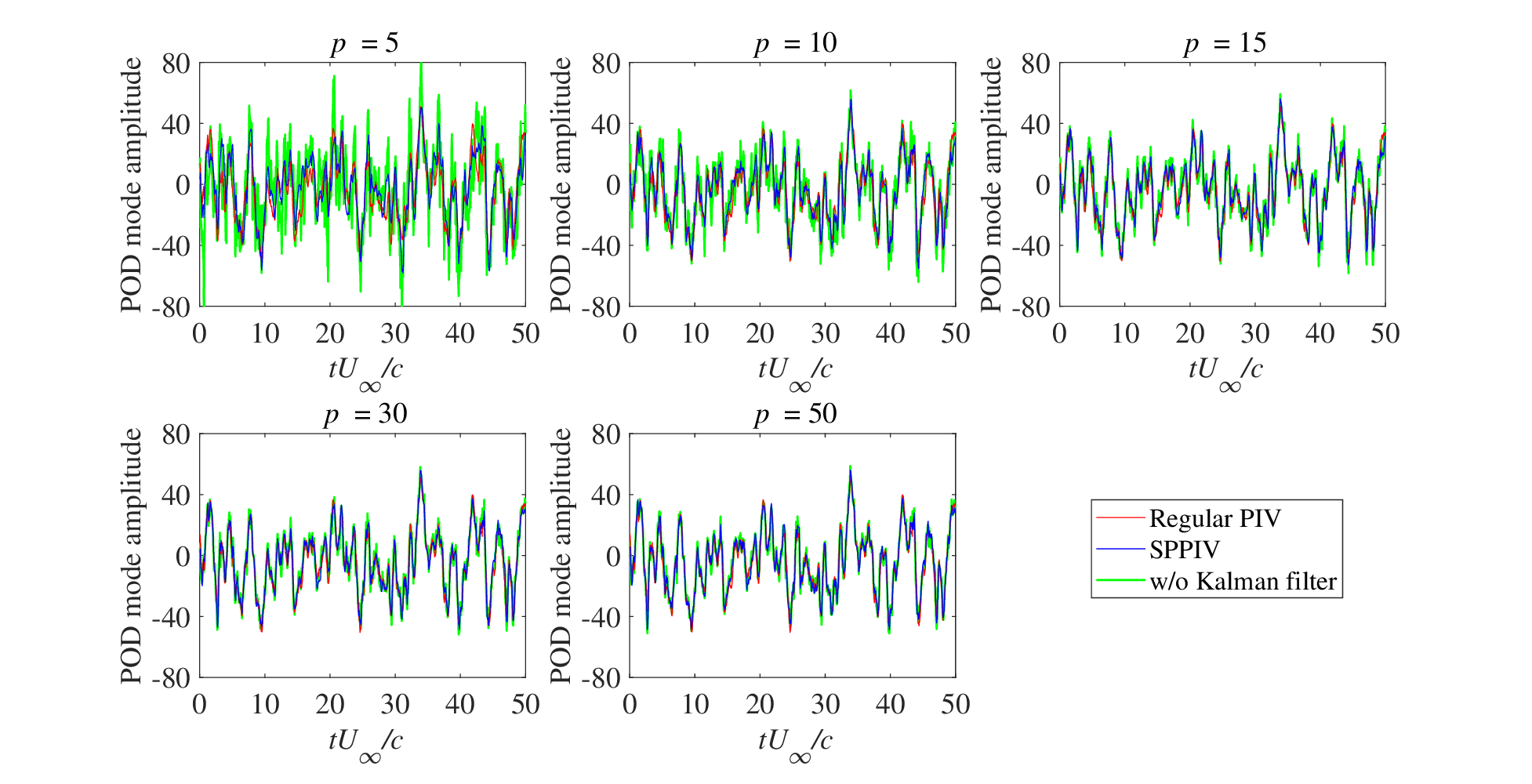}
    \caption{\reviewerB{The time histories of the first mode estimated w/ and w/o the Kalman filter in the SPPIV framework ($r$ = 10)}}
    \label{fig:timehistoriesofPODmodes_r10_woKF}
\end{figure}

\begin{figure}
    \centering
    \includegraphics[width=11.5cm]{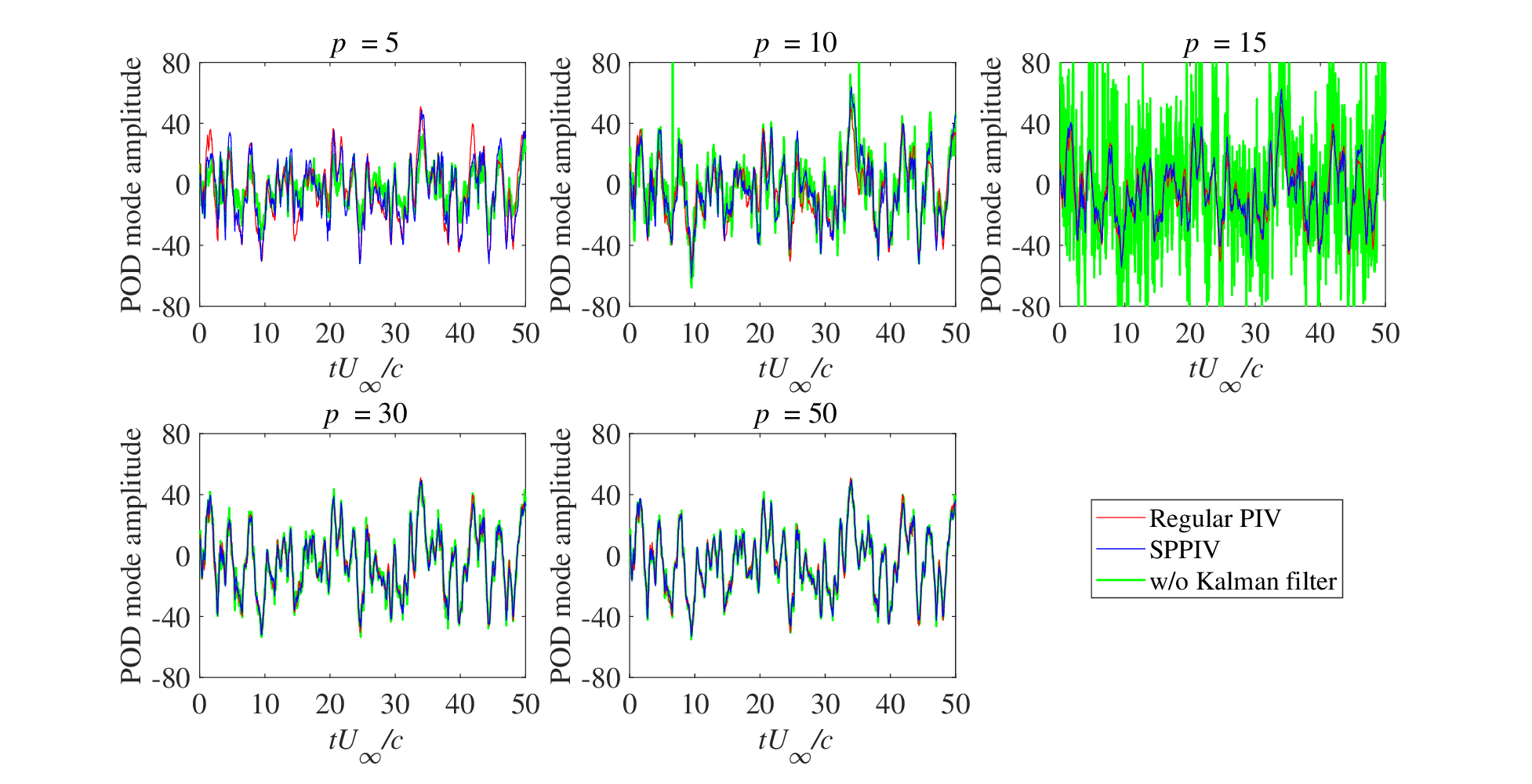}
    \caption{\reviewerB{The time histories of the first mode estimated w/ and w/o the Kalman filter in the SPPIV framework ($r$ = 30)}}
    \label{fig:timehistoriesofPODmodes_r30_woKF}
\end{figure}

\begin{figure}
    \centering
    \includegraphics[width=11cm]{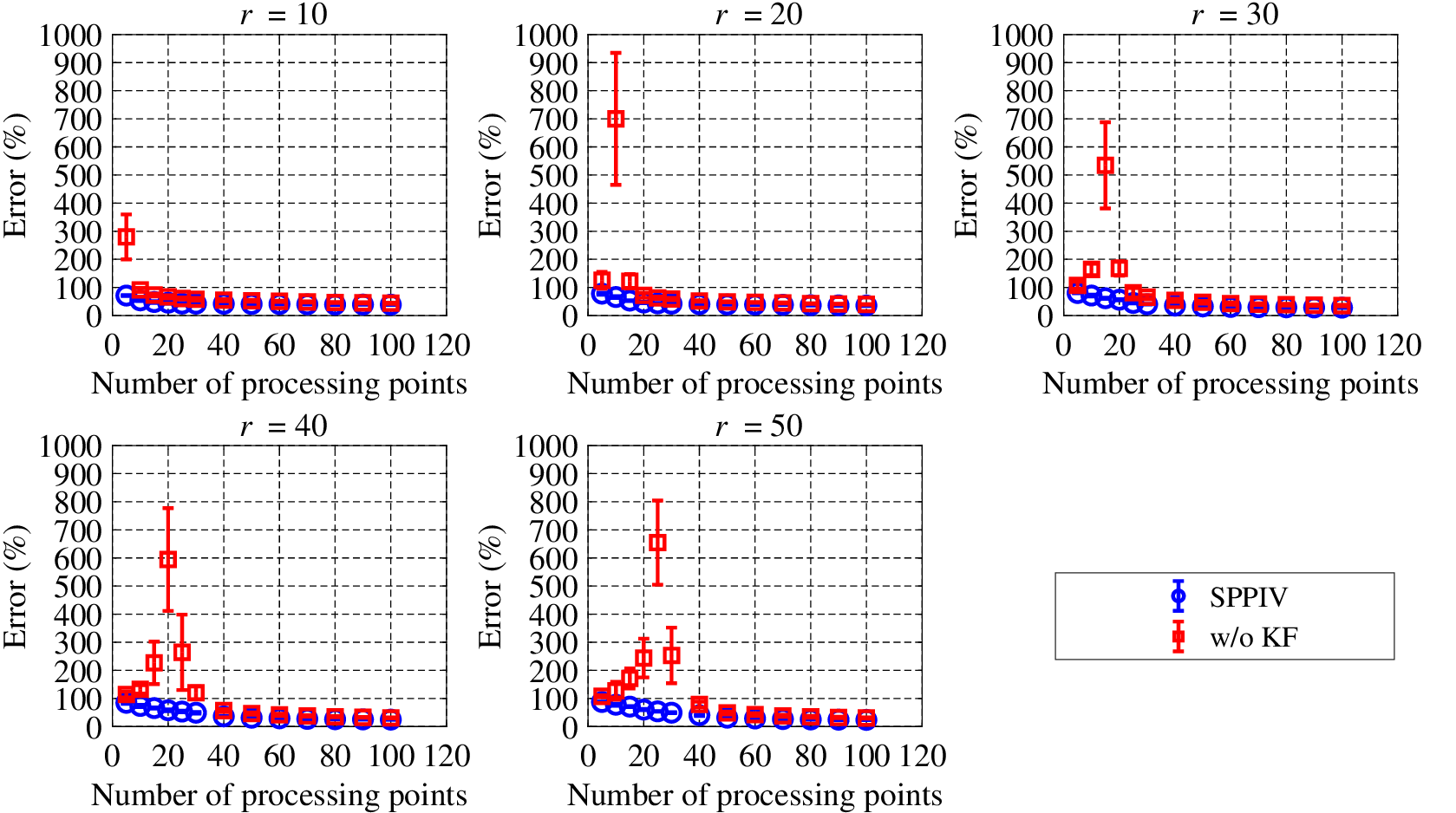}
    \caption{Comparing the error of the estimation w/ and w/o the Kalman filter ($p =$ 5 - 100)}
    \label{fig:error_5-100_woKF}
\end{figure}

\begin{figure}
    \centering
    \includegraphics[width=11cm]{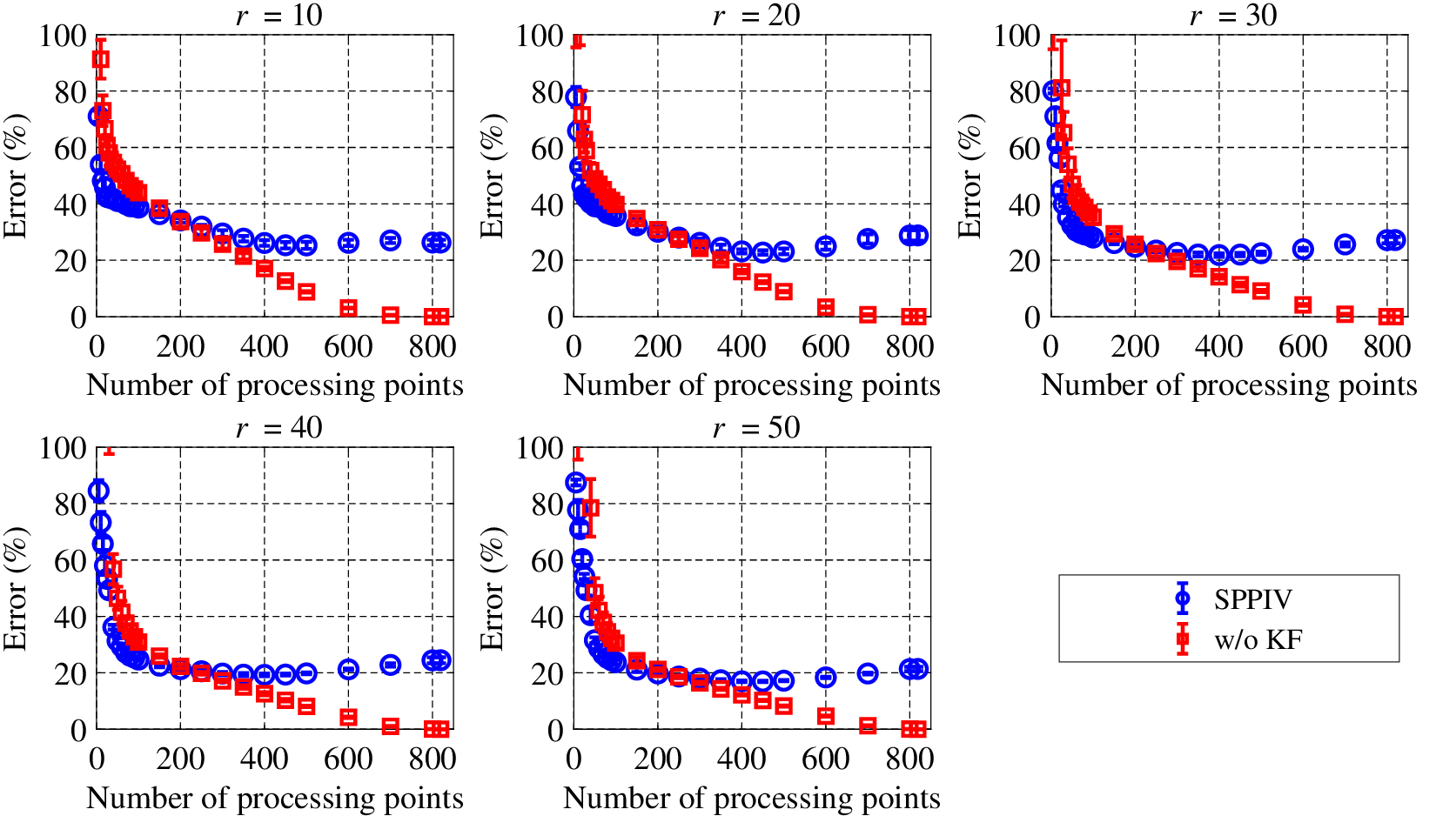}
    \caption{Comparing the error of the estimation w/ and w/o the Kalman filter ($p =$ 5 - 819)}
    \label{fig:error_5-819_woKF}
\end{figure}

Here, the estimation accuracy and the processing time per step are evaluated for the cases with and without the Kalman filter in the SPPIV framework, whereas the latter adopts the estimation only by the observation equation of the sparse flow velocity. Figures~\ref{fig:timehistoriesofPODmodes_r10_woKF} and \ref{fig:timehistoriesofPODmodes_r30_woKF} show the time histories of the representative POD mode estimated by each method in the case of $r=10$ and $r=30$, respectively. The error is calculated by Eq.~\ref{eq:mode_error_all} for these time histories. Figures~\ref{fig:error_5-100_woKF} and \ref{fig:error_5-819_woKF} show the errors when the number of processing points are 5 to 100 and when the number of processing points are 5 to 819, respectively. Here, the two plots with different ranges are employed and all the trends of errors are clearly shown because the error of the estimation without the Kalman filter is significantly large when the number of processing points is close to that of the number of modes. 


The flow velocities of two components in the freestream and vertical directions can be obtained from one correlation window (processing point) in PIV measurements, and therefore, the observation data of the same dimension as the number of modes can be obtained from $r/2$ processing points. Figure~\ref{fig:error_5-100_woKF} shows that the error is the largest at $p = r/2$  and the estimation accuracy is improved by increasing the number of processing points once after it becomes an overestimation system, when the Kalman filter is not applied. In addition, while estimation with the Kalman filter is more accurate for a smaller number of processing points, the estimation without the Kalman filter is more accurate when the number of processing points increases up to approximately 200 or more. \reviewerA{This shows that the linear model applied to SPPIV is not good at estimating higher-order modes, and the accuracy does not improve beyond a certain level even if the number of processing points increases. This might be because of the simplified noise model in the linear state-space model, and it might be improved by modifying both system and observation noise models. On the contrary, the pseudo inverse calculation might accurately recover the original field due to its simple situation without considering system noise if the number of processing points is sufficiently large. However, the main point of the present paper is demonstration of SPPIV which adopts the sparse processing points of less than 100, and therefore, the issues for the cases with a large number of sparse processing points are left for the future study. }

\begin{figure}
    \centering
    \includegraphics[width=11cm]{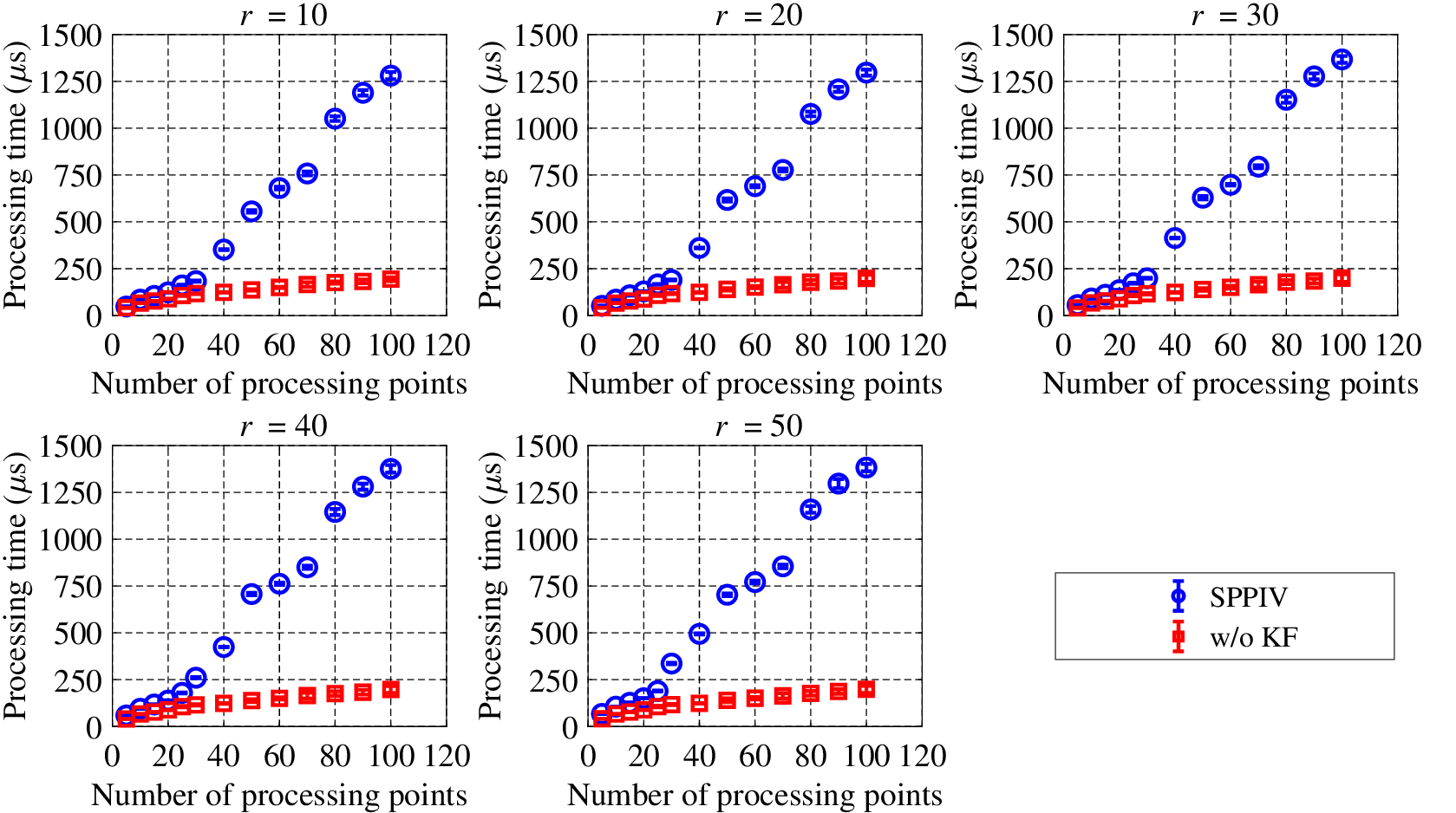}
    \caption{The processing time per step of the estimation w/ and w/o the Kalman filter}
    \label{fig:processingtime_woKF}
\end{figure}

Figure~\ref{fig:processingtime_woKF} shows the results of measuring the processing time per step in the case of estimation with and without the Kalman filter in the SPPIV framework. 
Figure~\ref{fig:processingtime_woKF} illustrates that the processing time per step is shorter in the case of estimating only the observation equation than that using the Kalman filter. However, Fig.~\ref{fig:error_5-100_woKF} shows that the advantage of improving the estimation accuracy by applying the Kalman filter is substantial when the number of processing points is small. Therefore, the application of Kalman filter is useful for estimation of the flow field from a small number of processing points in the SPPIV framework. 

\subsection{Comparison of results with mismatched training data}

In this section, we verify how much the estimation accuracy changes when the angles of attack in the training data and the test data are different. The angle of attack is set to be 14, 16, 18, 20, and 22~deg. The test data employs 1000 pairs of particle images of each angle of attack case, which is different from the training data. The number of processing points $p$ is 5, 10, 15, 20, 25, and 30 in the top 10 mode estimation.

\begin{figure}
    \centering
    \includegraphics[width=11cm]{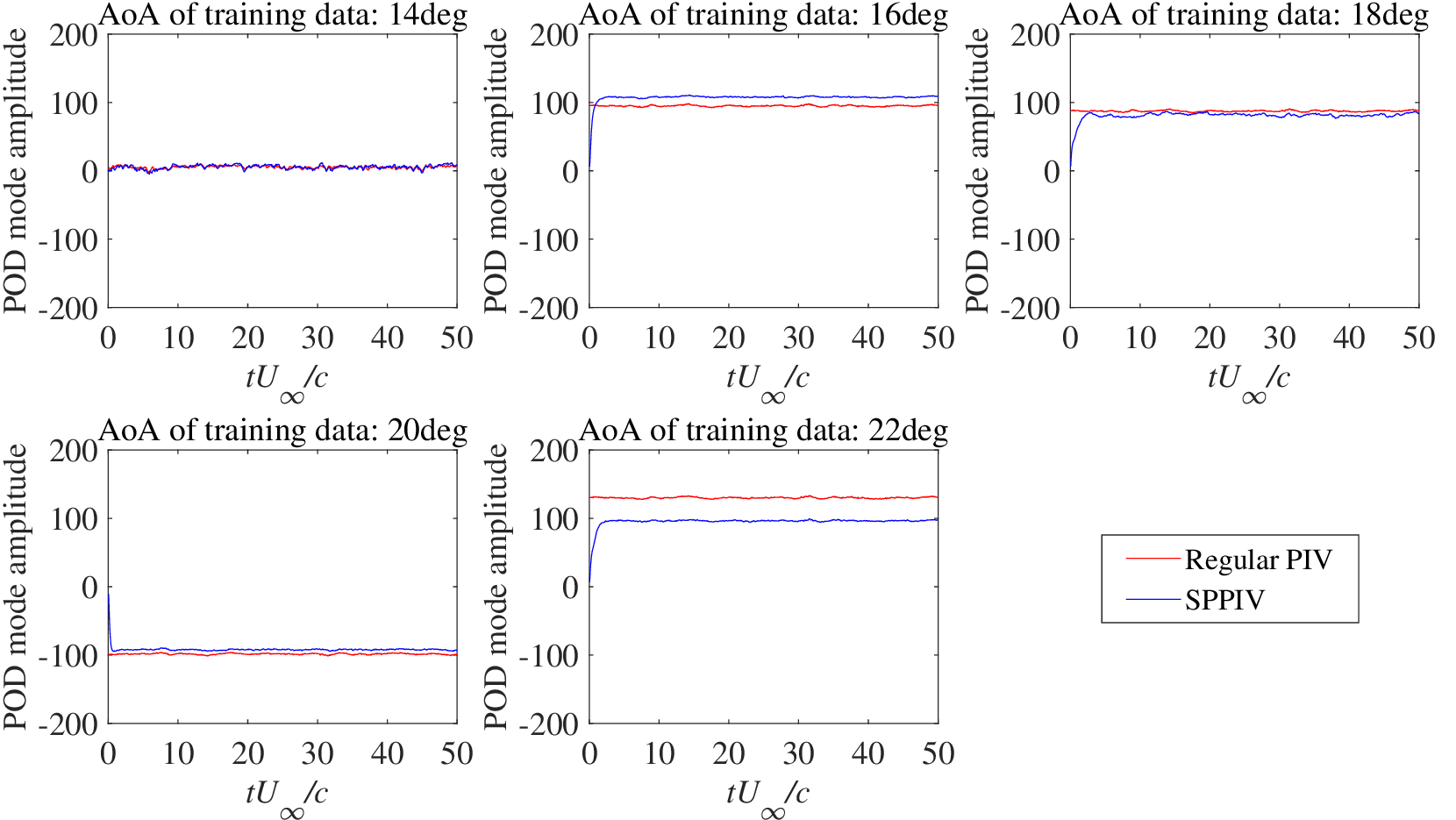}
    \caption{The time histories of the first mode when changing the angle of attack of training data (angle of attack of test data is 14~deg, $p = 5$, and $r = 10$)}
    \label{fig:timehistoriesofPODmodes_changetrain_14deg}
\end{figure}

\begin{figure}
    \centering
    \includegraphics[width=11cm]{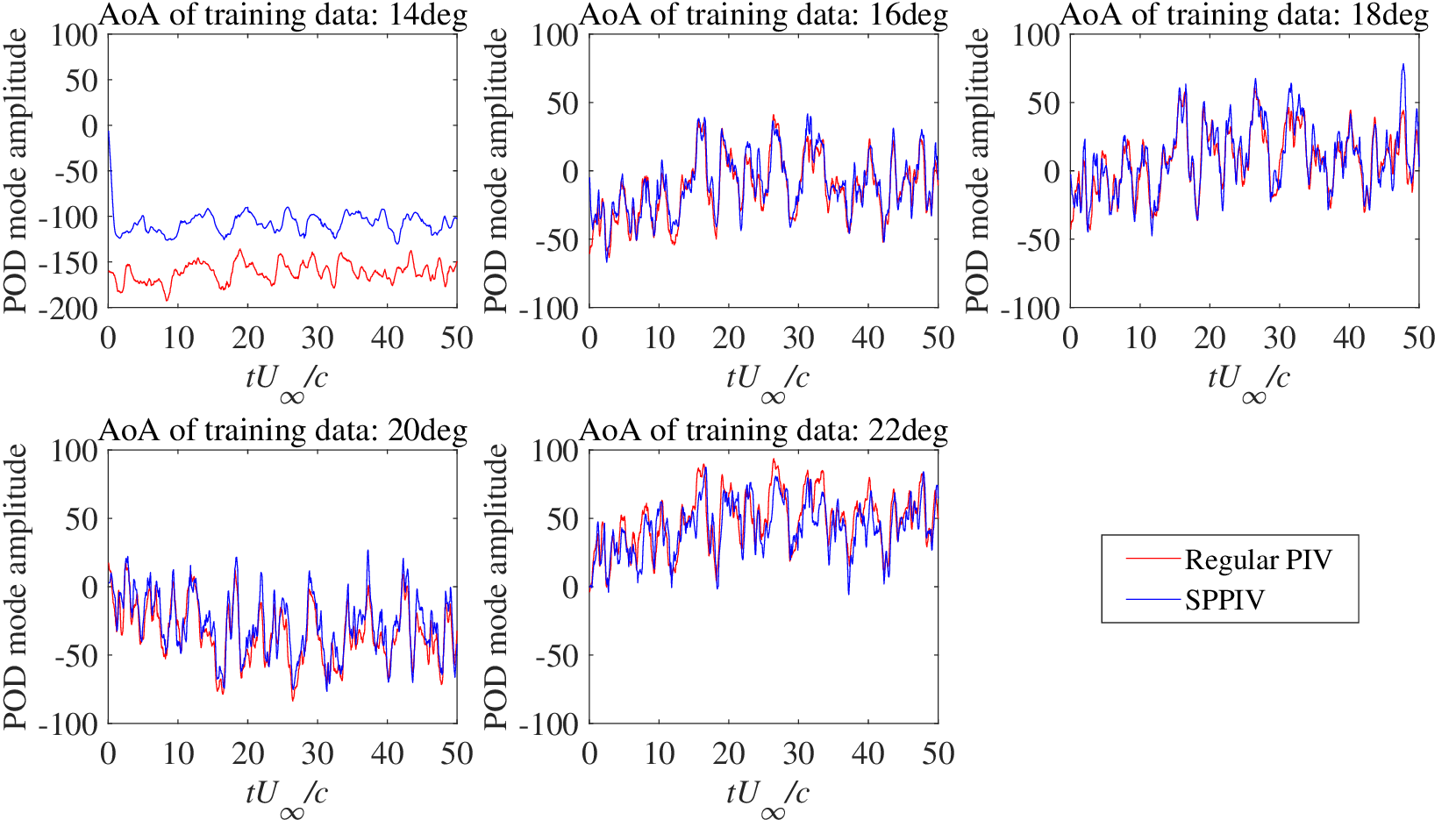}
    \caption{The time histories of the first mode when changing the angle of attack of training data (angle of attack of test data is 18~deg, $p = 5$, and $r = 10$)}
    \label{fig:timehistoriesofPODmodes_changetrain_18deg}
\end{figure}

\begin{figure}
    \centering
    \includegraphics[width=11cm]{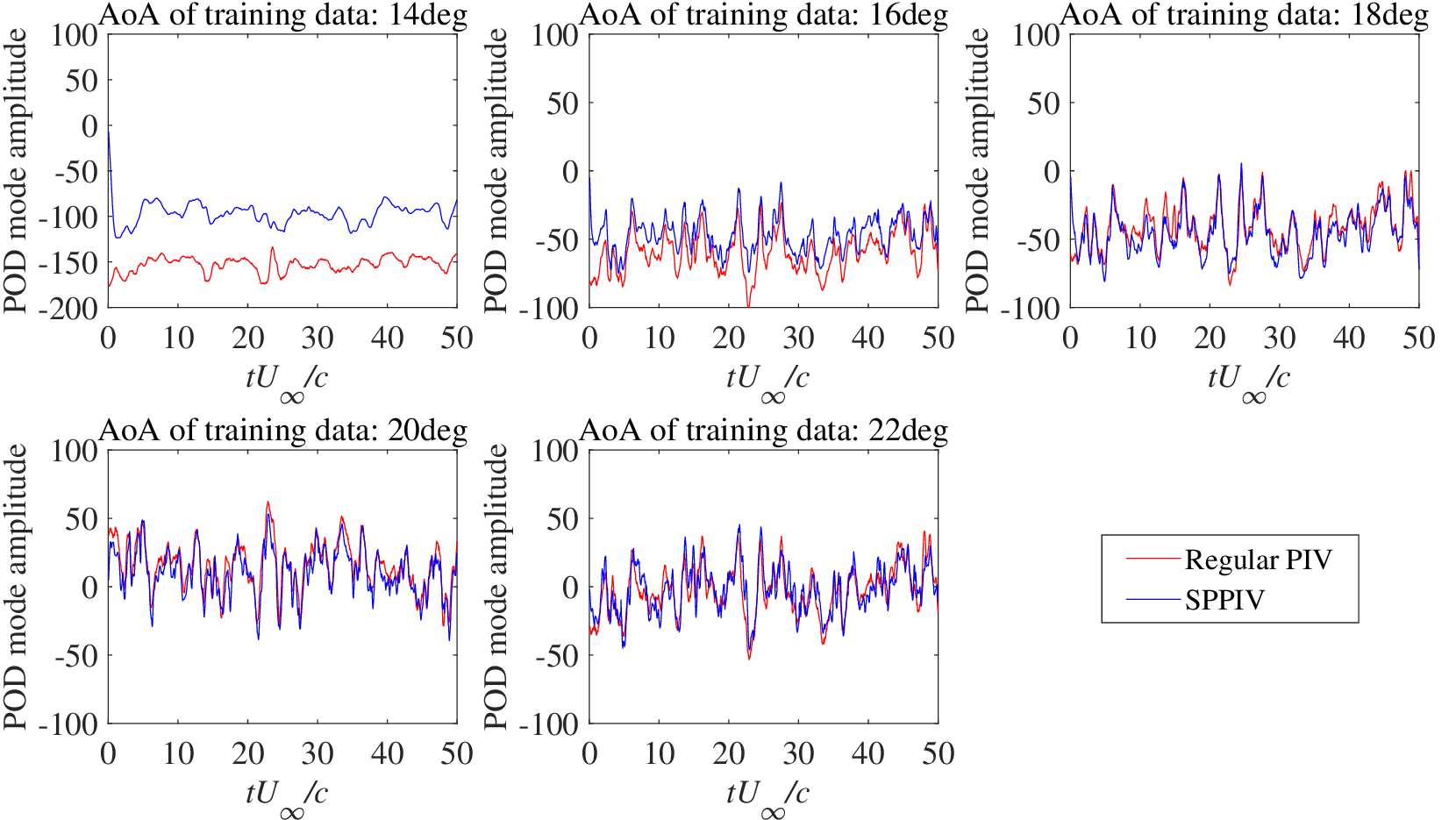}
    \caption{The time histories of the first mode when changing the angle of attack of training data (angle of attack of test data is 22~deg, $p = 5$, and $r = 10$)}
    \label{fig:timehistoriesofPODmodes_changetrain_22deg}
\end{figure}

\begin{figure}
    \centering
    \includegraphics[width=11cm]{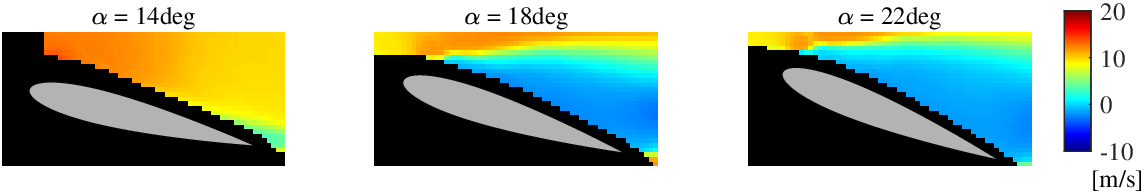}
    \caption{The time-averaged flow velocity fields of freestream direction of training data at the angles of attack of 14, 18, and 22 deg}
   \label{fig:timeaverageflowfields_trainingdata}
\end{figure}

Figures~\ref{fig:timehistoriesofPODmodes_changetrain_14deg}, \ref{fig:timehistoriesofPODmodes_changetrain_18deg}, and \ref{fig:timehistoriesofPODmodes_changetrain_22deg} show the time histories of the first mode estimated using training data with the angle of attack of 14 to 22~deg when the angle of attack of test data is 14, 18, and 22~deg, respectively.
Figure.~\ref{fig:timehistoriesofPODmodes_changetrain_14deg} illustrates that a large offset is included in the first mode when the case with the angle of attack of 14~deg is estimated using the training data of another angle of attack. This is because the state of the flow field differs only at the angle of attack of 14~deg under this test condition,  and therefore, each mode is estimated so that the difference between the averaged flow velocity fields used for the training and test data could be compensated.  \reviewerB{Figure~\ref{fig:timeaverageflowfields_trainingdata} shows the time-averaged flow velocity fields at angles of attack of 14, 18, and 22~deg. The flow at the angle of attack of 14~deg was confirmed to be qualitatively different from that at the angle of attack of 16~deg or more in this test condition. The trailing-edge-separated flow was observed at the angle of attack of 14~deg
while the fully separated flow was observed at the angles of attack of 16~deg or more.}
Figures~\ref{fig:timehistoriesofPODmodes_changetrain_18deg} and \ref{fig:timehistoriesofPODmodes_changetrain_22deg} show that it is possible to estimate the fluctuation even if the angles of attack of the training data and the test data are different but the flow fields of the training data are similar to those in test data. On the other hand, it oscillates around 0 when the angles of attack of the training and test data are the same, but it does around the offset value when the angles of attack of the training and test data are different.
This is because each mode is estimated so that the difference between the average flow velocity fields used for the training and test data could be compensated, as in the case of the angle of attack of 14~deg. 

\begin{figure}
    \centering
    \includegraphics[width=11cm]{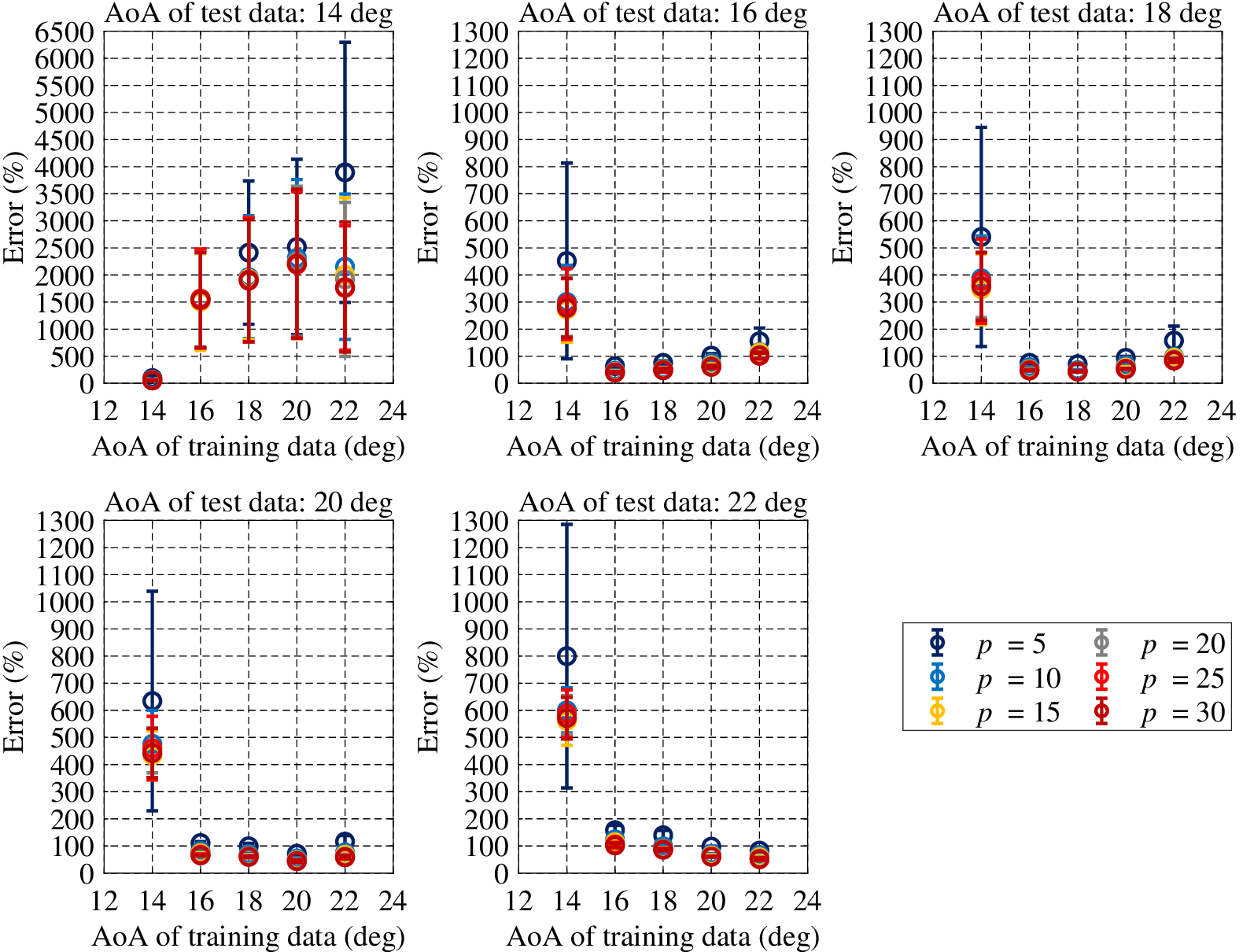}
    \caption{Comparing the error of SPPIV when changing the angle of attack of training data}
    \label{fig:plot_changetrain}
\end{figure}

\begin{figure}
    \centering
    \includegraphics[width=11cm]{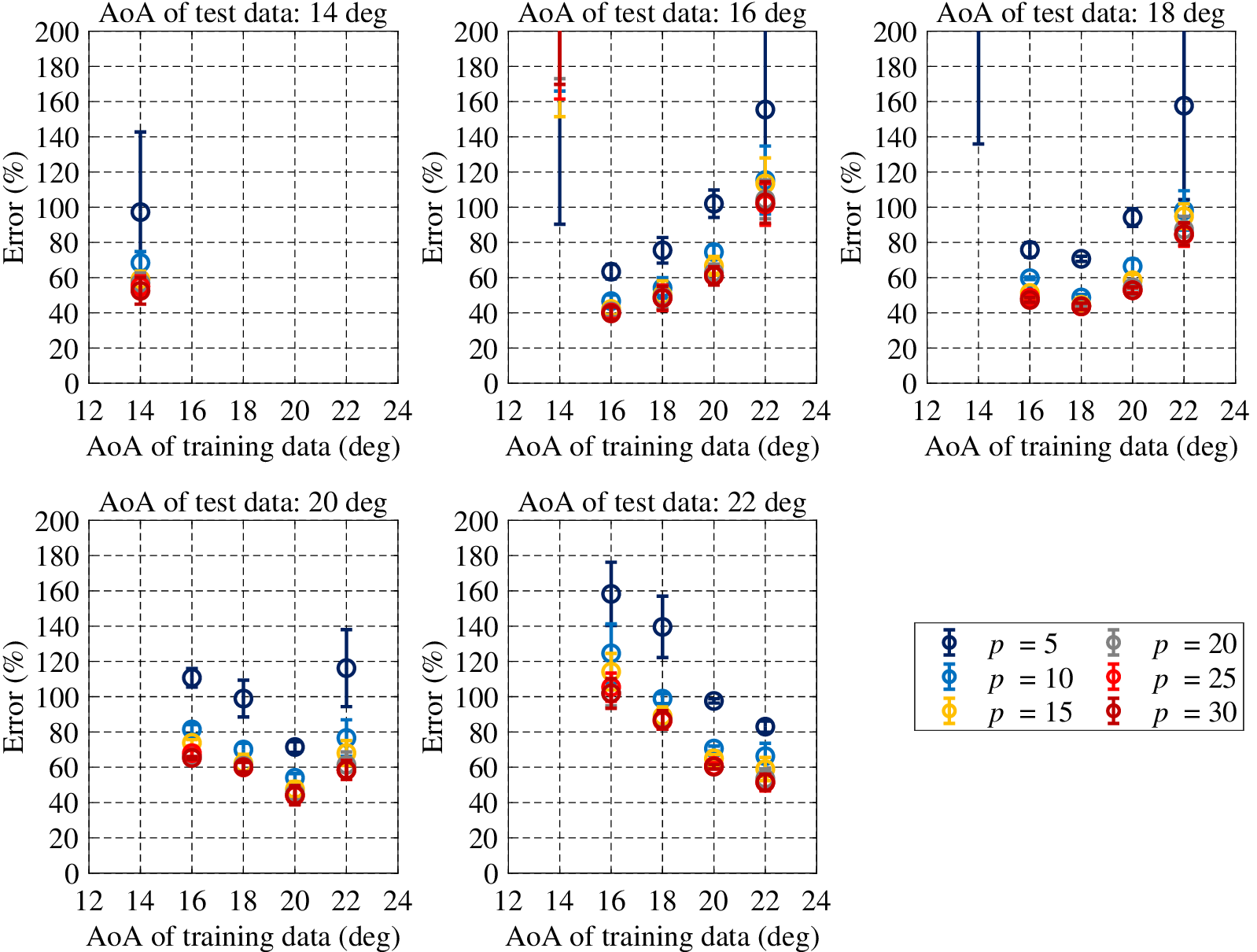}
    \caption{Comparing the error of SPPIV when changing the angle of attack of training data}
    \label{fig:plot_changetrain_2}
\end{figure}

The error is calculated and quantitatively evaluated for this POD mode. In the evaluation by the formula used in the previous section, the error may be underestimated due to the influence of the offset. The error normalized by the Frobenius norm of the POD mode variation is newly defined in Eq.~\ref{eq:mode_error_all_with_aveOrigin}, and the error calculated using this equation is discussed. Figures~\ref{fig:plot_changetrain} and \ref{fig:plot_changetrain_2} show the errors calculated by the formula. In Fig.~\ref{fig:plot_changetrain}, the range of the vertical axis is fixed for the case of the test-data angle of attack of 16 deg or more, while the range of the vertical axis is different for the case of 14 deg because of the larger error than those in other cases. Here, the two plots with different ranges are employed and the trends of errors are clearly shown.

\begin{align}
\label{eq:mode_error_all_with_aveOrigin}
    \epsilon &= \frac{\sqrt{\sum_{i=1}^{r}\sum_{j=1}^{m}(Z_{i,j}^{\rm{ref}}-\hat{Z}_{i,j})^2}}{\sqrt{\sum_{i=1}^{r}\sum_{j=1}^{m}{(Z_{i,j}^{\rm{ref}} - \overline{Z_{i}^{\rm{ref}}})^2}}}
\end{align}

Figure~\ref{fig:plot_changetrain} shows that flow fields at the angle of attack of 14~deg cannot be estimated from the training data of other angles of attack. \reviewerB{This is because the flow field is qualitatively different between the test and training data, as discussed for Figs.~\ref{fig:timehistoriesofPODmodes_changetrain_14deg}-\ref{fig:timeaverageflowfields_trainingdata}. The transition to the fully separated flow is considered to occur at between 14~deg and 16~deg in this test condition. Therefore, training data should be at least qualitatively the same type of flow fields, which implicates that the training data at the same angle of attack as the test data should be used in the condition around the transition. On the other hand, }
Fig.~\ref{fig:plot_changetrain_2} illustrates that the flow field can be estimated with the 20\%-error increase compared with the same angle-of-attack case when the angle of attack is 16~deg or more at which the flow is separated and the angle of attack difference between the test and training data is approximately $\pm2$~deg. In addition, the error in the case in which the angle of attack difference between the test and training data is $\pm4$~deg does not become so large when the number of processing points is 10 or more, which corresponds to the overestimated system.  
Many of the processing point positions obtained in this study are basically located near the shear layer. Fluctuations in the shear layer can be observed at the optimized processing points obtained from the training data, when the difference in the angles of attack is approximately $2$~deg.   On the other hand, the fluctuation near the shear layer cannot be measured when the angle of attack difference between the test and training data is larger than $2$~deg.  Therefore,  it is recommended to employ training data for each angle of attack in flow field estimation using SPPIV. \reviewerB{However, if that is difficult, the training data of similar angle of attack should be used when the flow field of the training data is similar to that of the test data. Since it is possible to estimate the flow field with less-than-20\% increase in the error within the difference of approximately $\pm2$~deg when the angle of attack is 16~deg or more, it is reasonable to acquire the training data at least every 4~deg after the transition to fully separated flow.} 

\subsection{\reviewerB{Recommendation for SPPIV setup}}
\reviewerB{In this section, we overview the results discussed in previous sections and describe the recommendations of the experimental conditions for the accurate estimation of the flow field using SPPIV.}

\reviewerB{The recommendation on the number of processing points is described below. The number of processing points needs to increase for the improvement of the estimation accuracy. However, there is a trade-off between the estimation accuracy and the processing time. The real-time measurement would be failed as will be discussed in the next section, when the processing time per step exceeds the image pair acquisition cycle. Therefore, the number of processing points is recommended to increase within the range the processing time is less than that required for the real-time measurement. In this study, the number of processing points of 20 was confirmed to improve the estimation accuracy most in the real-time observation. }

\reviewerB{With regard to the number of modes, it is not recommended to increase the number of modes too much since the estimation error using the number of modes of 10 to 30 becomes the lowest when the number of processing points is less than 30. On the other hand, the number of processing points could be increased to 50-100 in a real-time measurement when the slower flow is estimated or a higher-performance computer is utilized. In that case, the large number of modes such as 50 would achieve the lowest estimation error. Therefore, the number of modes needs to increase depending on the number of processing points.}

\reviewerB{
With regard to the estimation methods in SPPIV, it is recommended to use the Kalman filter especially when the number of processing points are small. 
}

\reviewerB{
Finally, with regard to the training data for SPPIV, it is recommended to employ the training data for each angle of attack in flow field estimation using SPPIV. On the other hand, when the flow field does not change qualitatively depending on the angle of attack, such as after the transition from the trailing-edge-separated flow to fully separated flow, it is possible to estimate the flow field with less than the 20\%-increased error within the difference of approximately $\pm2$~deg. Therefore, it is necessary to acquire the training data for each angle of attack around the transition condition, and it is reasonable to acquire the training data at least every 4~deg after the transition.}

\section{Results and Discussions of Real-Time Measurement}
\begin{figure}
    \centering
    \includegraphics[width=11cm]{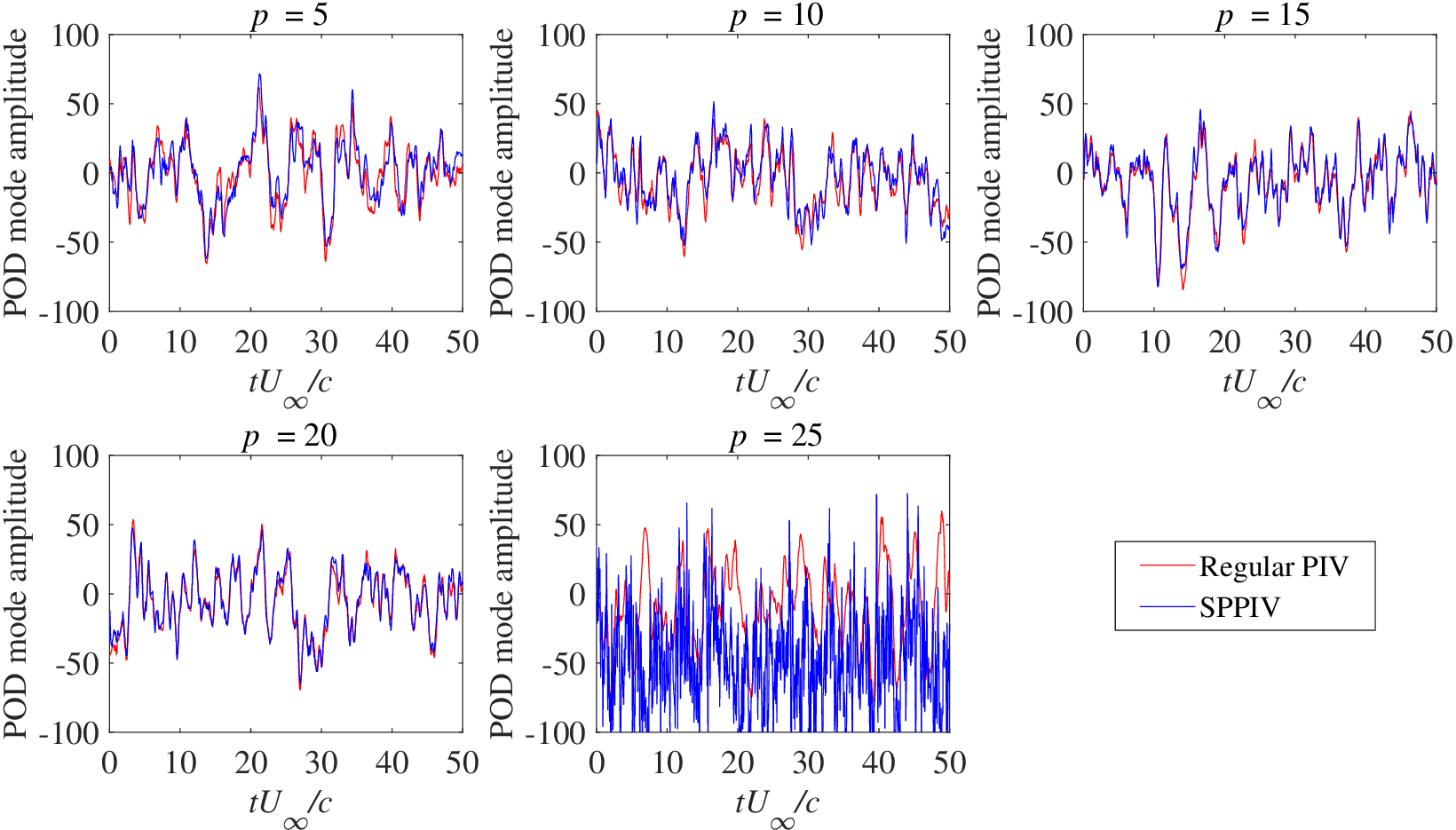}
    \caption{Time histories of the first mode obtained from real-time measurement}
    \label{fig:PODmode_RealtimeSPPIV}
\end{figure}

\begin{figure}
    \centering
    \includegraphics[width=11cm]{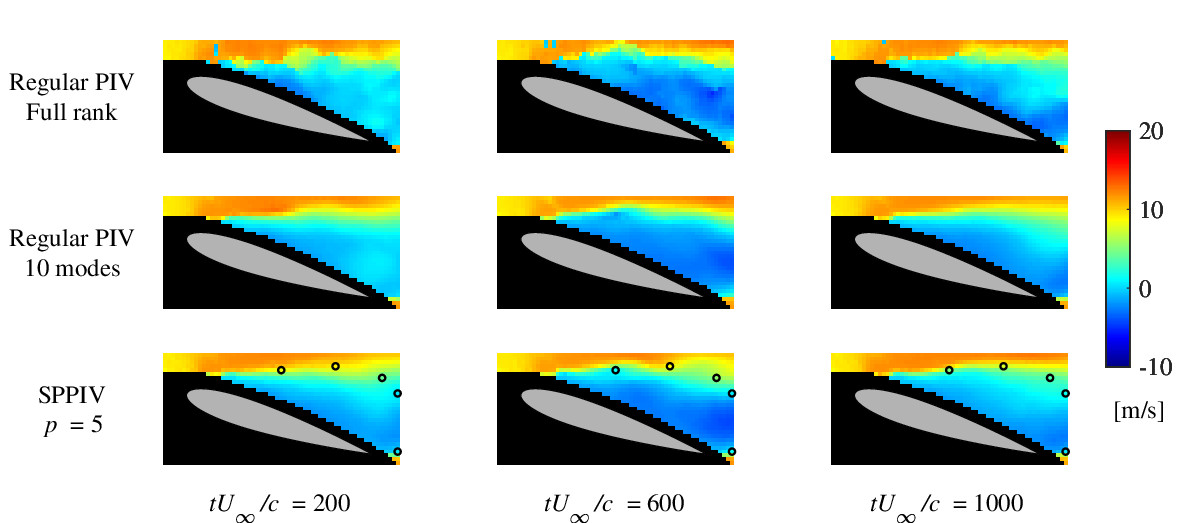}
    \caption{The freestream direction velocity fields reconstructed by POD modes obtained from real-time measurement ($r = 10$, $p = 5$) Here, circles show the sparse processing interrogation windows.}
    \label{fig:Velocityfield_RealtimeSPPIV}
\end{figure}

\begin{figure}
    \centering
    \includegraphics[width=8cm]{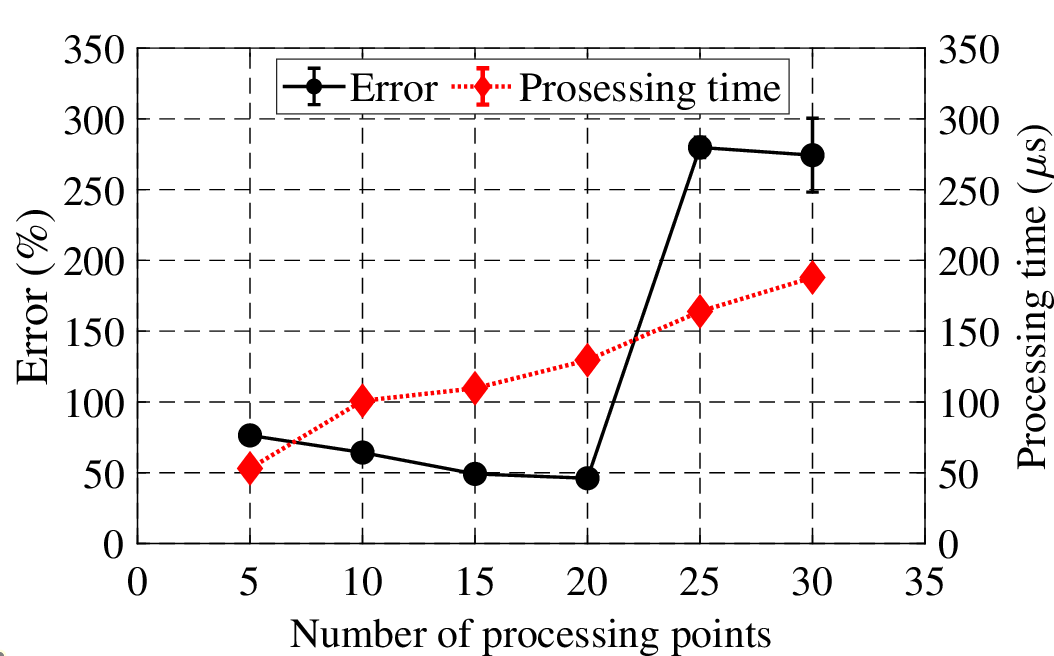}
    \caption{\reviewerB{The estimation error and the processing time per step of real-time SPPIV}}
    \label{fig:plot_RealtimeSPPIV}
\end{figure}

In this study, real-time measurement was performed at an angle of attack of 18~deg and the freestream velocity of 10~m/s. Figure~\ref{fig:PODmode_RealtimeSPPIV} shows the time histories of the first mode acquired by the real-time measurement, and Fig.~\ref{fig:Velocityfield_RealtimeSPPIV} shows the reconstructed freestream component field of the flow velocity when $p = 5$. Figure~\ref{fig:plot_RealtimeSPPIV} presents the error obtained by Eq.~\ref{eq:mode_error_all} and the processing time per step in the real-time measurement. The plots of error and processing time show the average data of three time measurements, and the error bars represent the standard deviation. The low-dimensional data shown in Fig.~\ref{fig:Velocityfield_RealtimeSPPIV} was obtained by the same process adopted for the results shown in Fig.~\ref{fig:verocityfieldsu_direction}.

\begin{figure}
    \centering
    \includegraphics[width=8cm]{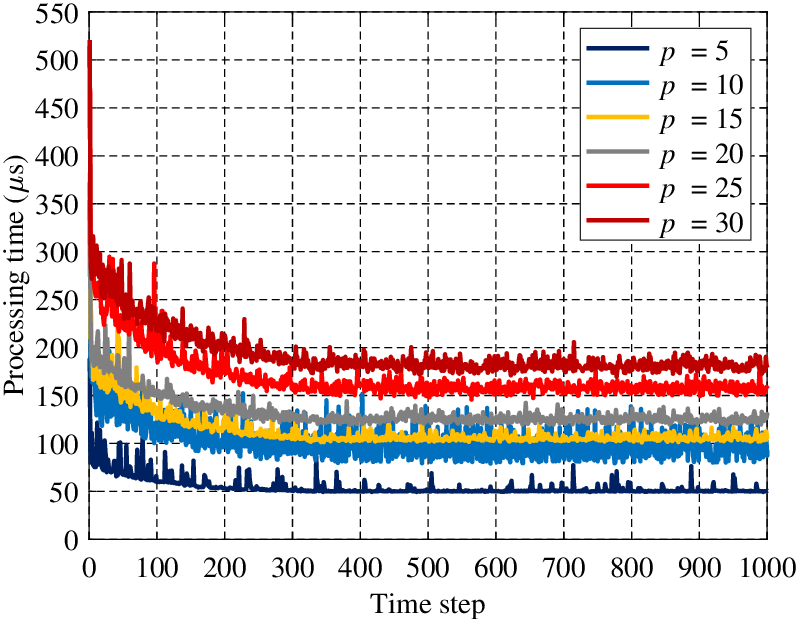}
    \caption{The time histories of processing time in real-time SPPIV}
    \label{fig:Timehistory_Processtime}
\end{figure}

Figure.~\ref{fig:plot_RealtimeSPPIV} shows that the error value increases when the number of processing points is 25 or more. Therefore, the real-time measurement could be realized under the test conditions in which the number of processing points is 20 or less. The failure of the real-time measurement seems to be because of the processing time and the naive implementation of the program code. Here, Fig.~\ref{fig:Timehistory_Processtime} shows the temporal change in the processing time per step. 

Figure~\ref{fig:plot_RealtimeSPPIV} illustrates that the average processing time per step can be suppressed to less than the image acquisition cycle even in the cases when the number of processing points is 25 or more. However, Fig.~\ref{fig:Timehistory_Processtime} shows that the processing time per step at the initial stage of measurement is long in the real-time measurements. When the number of processing points is 25 or more, the measurement was considered to fail because the processing with an appropriate image pair could not be performed due to the influence of the processing time per step at the initial steps. This problem seems to be possibly resolved by starting the measurement after the processing time per step becomes stably short, or by using a more customized program code. In addition, the application of the steady-state Kalman filter will help us accelerate the computation and relax this problem by reducing the calculation of the matrix inversion. 

\section{Conclusions}
In this paper, we overviewed, evaluated, and demonstrated SPPIV for real-time measurement of flow field. It should be noted that we succeeded in the real-time measurement of high-speed flow by sparse observation for the first time.

\reviewerB{First, the offline analyses were conducted and the characteristics of SPPIV were investigated.
    The flow velocity field can be estimated from a small number of processing points by applying SPPIV. The estimation accuracy is improved by increasing the number of processing points. Besides, it is improved by increasing the number of modes depending on the number of processing points. 
    However, the processing time per step increases in proportion to the number of processing points. 
    Therefore, it is necessary to set an optimal number of processing points. 
    Furthermore, the application of the Kalman filter significantly improved the estimation accuracy with a small number of processing points.
    In addition, the flow velocity fields with different angles of attack are used as the training data with that of test data. The estimation using SPPIV is found to be reasonable if the difference in the angle of attack between the training and test data is equal to or less than 2~deg and the flow phenomena of the training data are similar to that of the test data. Therefore, training data should be prepared at least every 4~deg. Note that the flow field of the training data is qualitatively different from that of the test data, such as the transition from the trailing-edge-separated flow to the fully separated flow, it is necessary to acquire the training data for each angle of attack.}

\reviewerB{Then, we successfully conducted the real-time measurement as the demonstration of SPPIV for the first time. The real-time measurement is found to be possible at a sampling rate of 2000~Hz at 20 or less processing points in the top 10 modes estimation as expected by the off-line analyses. }



\section*{Acknowledgments}
The present study was supported by JST FOREST(JPMJFR202C) and ACT-X (JPMJAX20AD) and KAKENHI (20H00279,21H04586), Japan. K.~Nakai was supported by JST CREST (JPMJCR1763). The authors are grateful to Prof. Yoshihiro Watanabe of Tokyo Institute of Technology who shared their experience in the experiment using real-time cameras in the early stage of the present study. 

\section*{Conflict of interest}
The authors declare that they have no conflict of interest or competing interests.

\bibliographystyle{spbasic}
\bibliography{xaerolab.bib}

\end{document}